\documentclass[a4paper,UKenglish,cleveref, autoref, thm-restate]{lipics-v2021}
\nolinenumbers
\hideLIPIcs  %uncomment to remove references to LIPIcs series (logo, DOI, ...), e.g. when preparing a pre-final version to be uploaded to arXiv or another public repository

\usepackage{graphicx}
\usepackage{algorithm}
\usepackage[noend]{algpseudocode}
\usepackage{amsmath}
\usepackage{multirow}
\usepackage{graphicx}
\usepackage{xcolor}
\usepackage{amsfonts}
\usepackage{amsthm}
\usepackage{pifont}
\usepackage{xspace}
\usepackage{mathtools}
\usepackage{enumitem}
\usepackage{lineno}
\usepackage{thmtools}
\usepackage{thm-restate}
\usepackage{relsize}
\usepackage{hhline}% http://ctan.org/pkg/hhline
\usepackage{etoolbox}
\usepackage{thmtools}
\usepackage{thm-restate}
\usepackage{cleveref}
\usepackage[most]{tcolorbox}
\usepackage{multicol}
\usepackage{changepage}
\usepackage{algpseudocode}
\usepackage{hhline}
\usepackage{etoolbox}
\usepackage{relsize}
\usepackage{bm}
\usepackage{scalefnt}
\usepackage{wrapfig}

\setlist[description]{leftmargin=*,labelindent=*}

\newcommand{\alglinenoNew}[1]{\newcounter{ALG@line@#1}}
\newcommand{\alglinenoPop}[1]{\setcounter{ALG@line}{\value{ALG@line@#1}}}
\newcommand{\alglinenoPush}[1]{\setcounter{ALG@line@#1}{\value{ALG@line}}}

\newcommand{\mypara}[1]{\smallskip\noindent\textbf{#1.}}

\newcommand{\ia}{\textit{i}}
\newcommand{\ib}{\textit{ii}}
\newcommand{\ic}{\textit{iii}}
\newcommand{\id}{\textit{iv}}
\newcommand{\iie}{\textit{v}}

\newcommand{\com}[1]{}

\algdef{SE}[Receiving]{Receiving}{EndReceiving}[1]{\textbf{upon
		receiving}\ #1\ \algorithmicdo}{\algorithmicend\ \textbf{}}%
\algtext*{EndReceiving}

\algdef{SE}[Upon]{Upon}{EndUpon}[1]{\textbf{upon}\ #1\ \algorithmicdo}{\algorithmicend\ \textbf{}}%
\algtext*{EndUpon}

\newcommand\StateX{\Statex\hspace{\algorithmicindent}}
\newcommand\StateXX{\StateX\hspace{\algorithmicindent}}
\algrenewcommand\textproc{}% Used to be \textsc

\newcommand{\temph}[1]{\textbf{#1}}

\makeatletter \let\sv@thm\@thm \def\@thm{\let\indent\relax\sv@thm} \makeatother

\newcommand{\calN}{\mathbb{N}}
\newcommand{\calA}{\mathcal{A}}

\newcommand{\calS}{\mathcal{S}}

\newcommand{\calB}{\mathcal{B}}
\newcommand{\calC}{\mathcal{C}}
\newcommand{\calX}{\mathcal{X}}
\newcommand{\calD}{\mathcal{D}}
\newcommand{\calF}{\mathcal{F}}
\newcommand{\calG}{\mathcal{G}}

\newcommand{\NN}{\mathbb{N}}
\newcommand{\PP}{P}

\crefname{table}{table}{tables}
\crefname{table}{Table}{Tables}
\crefname{algocf}{alg.}{algs.}
\crefname{algocf}{Alg.}{Algs.}
\crefname{figure}{Fig.}{Figs.}
\crefname{figure}{fig.}{figs.}
\crefname{claim}{claim}{claims}
\crefname{claim}{Claim}{Claims}
\crefformat{chapter}{\S#2#1#3}
\crefmultiformat{chapter}{\S\S#2#1#3}{and~#2#1#3}{, #2#1#3}{, and~#2#1#3}

\crefformat{section}{\S#2#1#3}
\crefmultiformat{section}{\S\S#2#1#3}{and~#2#1#3}{, #2#1#3}{, and~#2#1#3}

\usepackage{enumitem}
\setlist{nosep} % or \setlist{noitemsep} to leave space around whole list
\setlist{itemsep=1pt, topsep=3pt}

\newcommand{\CD}{Cordial Dissemination\xspace}

\newcommand{ \AD}{All-to-All Dissemination\xspace}

\newcommand{\GD}{Grassroots Dissemination\xspace}

\newcommand{\calCD}{\calC\calD}

\newcommand{\preceqGD}{\preceq_{\calG\calD}}
\newcommand{\preceqCD}{\preceq_{\calCD}}

\newcommand{\Id}{\textit{Id}}

\title{\smaller Grassroots Systems:\\ Concept, Examples, Implementation and Applications} %TODO Please add

\titlerunning{Grassroots Systems} %TODO optional, please use if title is longer than one line

\author{Ehud Shapiro}{Department of Computer Science and Applied Math, Weizmann Institute of Science,  Israel}{ehud.shapiro@weizmann.ac.il}{}{}

\authorrunning{Ehud Shapiro} %TODO mandatory. First: Use abbreviated first/middle names. Second (only in severe cases): Use first author plus 'et al.'

\Copyright{Ehud Shapiro} %TODO mandatory, please use full first names. LIPIcs license is "CC-BY";  http://creativecommons.org/licenses/by/3.0/

 %TODO mandatory; please add comma-separated list of keywords

\category{} %optional, e.g. invited paper

\relatedversion{} %optional, e.g. full version hosted on arXiv, HAL, or other respository/website
\acknowledgements{}%optional

\begin{document}

\maketitle

\begin{abstract}
Informally, a \emph{grassroots system} is a distributed system that can have multiple instances, independent of each other and of any global resources,  that can interoperate once interconnected.
More formally, in a grassroots system the set of all correct behaviors of a set of agents $P$ is strictly included in the set of the correct behaviors of $P$ when embedded within a larger set of agents $P' \supset P$:  Included, meaning that members of $P$ are correct to ignore outside agents indefinitely; strictly included, meaning that the interaction of members of $P$ with outside agents is also a possible and correct behavior.

Client-server/cloud computing systems are not grassroots, and neither are systems designed to have a single global instance (Bitcoin/Ethereum with hardwired seed miners/bootnodes), and systems that rely on a single global data structure (IPFS, DHTs).
An example grassroots system would be a serverless smartphone-based  social network supporting multiple independently-budding communities that can merge when a member of one community becomes also a member of another.

Grassroots applications are potentially important as they may allow people to conduct their social, economic, civic, and political lives in the digital realm solely using the networked computing devices they own and operate (e.g., smartphones), free of third-party control, surveillance, manipulation, coercion, or rent seeking (e.g., by global digital platforms such as Facebook or Bitcoin).

Here, we formalize the notion of grassroots systems and grassroots implementations; specify an abstract grassroots dissemination protocol; describe and prove an implementation of grassroots dissemination for the model of asynchrony; extend the implementation to mobile (address-changing) devices that communicate via an unreliable network (e.g. smartphones using UDP); and discuss  how grassroots dissemination can realize applications that support digital sovereignty -- grassroots social networking and grassroots currencies.
The mathematical construction employs distributed multiagent transition systems to define the notions of  grassroots protocols and grassroots implementations,  to specify grassroots dissemination protocols and their implementation, and to prove their correctness.  The implementation uses the blocklace -- a distributed, partially-ordered generalization of the replicated, totally-ordered blockchain.
\end{abstract}

\section{Introduction}

\mypara{Motivation} Today, client-server/cloud computing is the dominant architecture in the digital realm,  supporting the global digital platforms we inhabit on a daily basis.  These platforms have adopted surveillance-capitalism~\cite{zuboff2019age} as their business model, which drives them to monitor, induce, and manipulate their inhabitants for profit.  Authoritarian regimes may have an added ingredient: A ``back-door'' to the global platform for the regime to monitor, censor, control, and even punish the digital behavior of its citizens.  

Blockchain technology and cryptocurrencies~\cite{bitcoin,buterin2014next} offer a radically-different architecture that is peer-to-peer and can be `permissionless',  open to participation by everyone. Still, a `peer' in this architecture is typically a high-performance server or, better yet, a high-performance server-cluster.   Despite the promise of openness and distribution of power, leading cryptocurrencies are also global platforms that are dominated---even controlled---by a handful of ultra-high-performance server-clusters~\cite{gencer2018decentralization}.  The operators of these server clusters are remunerated for their efforts via inflationary coin minting or transaction fees/gas paid by the platform users.
An alternative server-based architecture that strives for better distribution of power is federation, employed for example by BitTorrent~\cite{cohen2003incentives}, IPFS~\cite{benet2014ipfs}, and Mastodon~\cite{raman2019challenges} and peer-to-peer architectures such as distributed pub/sub systems~\cite{chockler2007constructing,chockler2007spidercast}, Gryphone~\cite{strom1998gryphon}, PeerSON~\cite{buchegger2009peerson}, Scuttlebutt~\cite{tarr2019secure}, and more.

Today's smartphones have orders-of-magnitude more computing power, memory, and network speed compared to the Unix workstations that were the workhorses of the Internet revolution.  Yet, they function mostly as adorned gateways to global digital platforms, with even peer-to-peer streaming being initiated and controlled by the cloud.  We believe that the lack of a business model that can incentivize `genuine' peer-to-peer applications and the lack of suitable architectural foundations for such applications are the culprit. 

Here, we are concerned with providing such architectural foundations, referred to as a grassroots architecture, a notion introduced and formally defined below.
Informally, a distributed system is \emph{grassroots} if it can have autonomous, independently-deployed instances---geographically and over time---that can interoperate once interconnected.  
A straightforward manifestation of a grassroots  system would be a peer-to-peer, serverless, smartphone-based social network that can form independently at multiple communities and at different times, and interoperate if and when the communities are interconnected, for example via a person that becomes a member of two hitherto-disjoint communities, similarly to Scuttlebutt~\cite{tarr2019secure}.

The grassroots dissemination protocol developed here can support such an applications~\cite{shapiro2023gsn}.
Grassroots applications are potentially important as they may allow people to conduct their social, economic, civic, and political lives in the digital realm solely using the networked computing devices they own and operate (e.g., smartphones), free of third-party control, surveillance, manipulation, coercion, or value-extraction (e.g., by global digital platforms such as Facebook or Bitcoin).

In general, a system designed to have a single global instance is not grassroots.
Client-server/cloud systems in which two instances cannot co-exist due to competition/conflict on shared resources (e.g., same web address), or cannot interoperate when interconnected, 
are not grassroots.
Neither are peer-to-peer systems that require all-to-all dissemination, including mainstream cryptocurrencies and standard consensus protocols~\cite{castro1999practical,yin2019hotstuff,keidar2021need}, 
since a community placed in a larger context cannot ignore members of the larger context.
Neither are systems that use a global shared data-structure such as pub/sub systems~\cite{chockler2007constructing,chockler2007spidercast}, IPFS~\cite{benet2014ipfs}, and distributed hash tables~\cite{stoica2001chord},  since a community placed in a larger context cannot ignore updates to the shared resource by others. 
While we do not prove this formally, federated systems such as BitTorrent~\cite{cohen2003incentives} with private trackers  and Mastodon~\cite{raman2019challenges},  in which federation is optional and there are no shared global resources, as well as Scuttlebutt~\cite{tarr2019secure}, may be grassroots.

\mypara{Contributions} While the notion of `grassroots' has intuitive appeal and is well-established in the social and political arena~\cite{castells1983city}, it has not received, to the best of our knowledge, any formal treatment.
One contribution of this work is its formal definition and characterization in the context of multiagent distributed systems.

Another contribution is a specification and a proven implementation for the model of asynchrony of perhaps the first grassroots dissemination protocol. The result is powerful enough to realize serverless social networking~\cite{shapiro2023gsn} and grassroots currencies~\cite{shapiro2024gc,lewis2023grassroots}.  The protocol is specified by the rather-abstract \GD protocol and implemented by the blocklace-based  \CD protocol, both specified as asynchronous distributed multiagent transitions systems~\cite{shapiro2021multiagent}, with pseudocode realizing the latter for the model of asynchrony and for mobile devices using unreliable communication (UDP).

\mypara{Related work} The blocklace is a distributed, partially-ordered counterpart to the replicated, totally-ordered blockchain, which functions as a fault-resilient, conflict-free replicated data type~\cite{almeida2024blocklace}.  It has already been employed to realize consensus protocols~\cite{keidar2023cordial}, grassroots social networking~\cite{shapiro2023gsn}, a supermajority-based payment system~\cite{lewispye2023flash} and a grassroots payment system~\cite{lewispye2023flash}.
All-to-all dissemination has been explored extensively over the past decades, under the notions of gossip~\cite{shah2009gossip} and epidemic~\cite{pastor2015epidemic} protocols. Here, we use all-to-all dissemination as a stepping stone to grassroots dissemination. 

The grassroots dissemination protocol must give credit to the  pioneering 1990's
Bayou project at Xerox PARC~\cite{demers1994bayou}.  It is also
reminiscent of the pub/sub model~\cite{chockler2007constructing,chockler2007spidercast}, where every agent is viewed as a publisher and every follower as a subscriber. While peer-to-peer protocols for the pub/sub model were developed (see~\cite{setty2012poldercast} for a review), to the best of our understanding they are not grassroots.  For example, the Poldercast protocol~\cite{setty2012poldercast} is not grassroots since
it assumes a global  set of topics that is disjoint from the set of agents participating in the protocol.  In particular, in Poldercast any infinite correct run of a group of agents $P$ that subscribes to a topic $t$, when repeated within a larger group $P'$ with members that publish to the same topic $t$, would violate liveness. The reason is that  members of $P$ that subscribe to $t$, when embedded within $P'$, should receive in such a run posts to $t$ by published by agents in $P'\setminus P$; but in such a run they do not.  In addition, there are social networks such as PeerSoN~\cite{buchegger2009peerson} that are peer-to-peer, but to the best of our understanding are not grassroots, for example because they employ Distributed Hash Tables~\cite{stoica2001chord} as a shared global resource.

\begin{wrapfigure}{r}{5.5cm}
\begin{center}
\includegraphics[width=5.5cm]{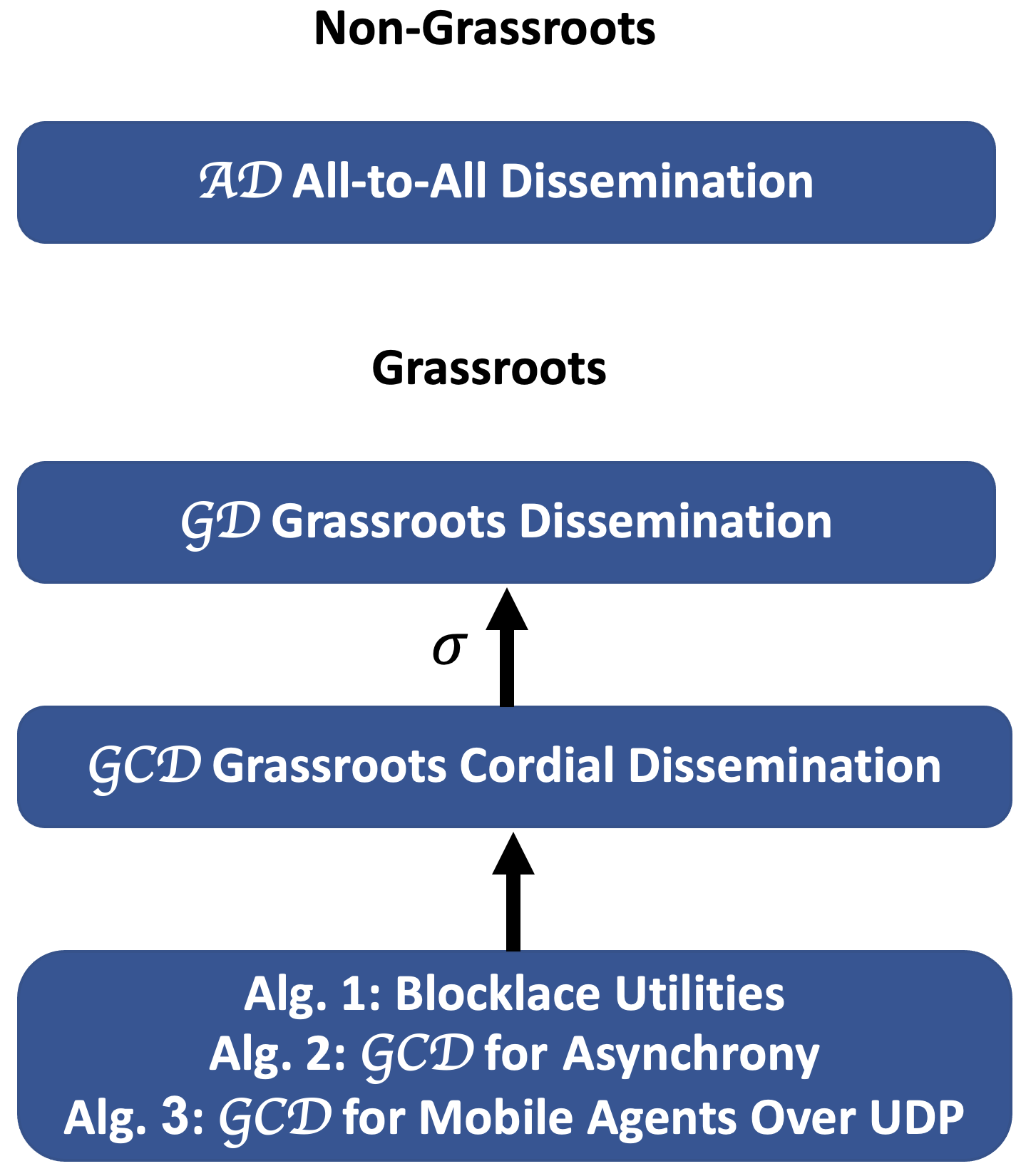}
\end{center}
\caption{Protocols and Implementations presented in this paper.
\AD  $\calA\calD$ (Def. \ref{definition:AD}), \GD  $\calG\calD$ (Def. \ref{definition:GD}),
 \CD  $\calCD$ (Def. \ref{definition:protocol-CD}), and the implementation  $\sigma$ (Def. \ref{definition:sigma-CD-GD}, Thm. \ref{theorem:CD-implements-GD}) of $\calG\calD$  by $\calCD$. 
Algorithms \ref{alg:blocklace}, \ref{alg:CD}, and  \ref{alg:CD-UDP} that implement  $\calCD$ for Asynchrony and UDP are presented as pseudocode.}
\label{figure:GD-stack}
\end{wrapfigure}
\mypara{Outline} The dissemination protocols introduced in this paper and their implementations are summarized in Figure \ref{figure:GD-stack}. The rest of the paper is organized as follows. Section \ref{section:grassroots} briefly introduces asynchronous distributed multiagent transition systems (expanded upon in Appendix \ref{section:preliminaries}) and implementations among them,  providing the mathematical basis for
defining grassroots protocols and grassroots implementations.  It then 
introduces the rather-abstract Grassroots Dissemination protocol $\calG\calD$ and proves it to be grassroots.
Section \ref{section:GD-implementation} describes the implementation of grassroots dissemination using the blocklace -- a distributed, partially-ordered generalization of the blockchain datatype.  It introduces blocklace basics;  the blocklace-based Grassroots Cordial Dissemination protocol $\calCD$; presents an implementation $\sigma$ of $\calG\calD$ by
$\calCD$ and proves it correct; provides pseudocode implementation of $\calCD$ for the model of asynchrony; and extends the implementation to mobile devices using unreliable communication (UDP).
Section \ref{section:applications} discusses the applications of grassroots dissemination to serverless social networking and grassroots currencies.
Section \ref{section:conclusions} concludes the paper.
Due to length limitations, proofs and some auxiliary material are relegated to the Appendices.

\section{Grassroots Protocols}\label{section:grassroots}

\subsection{Definition}

Informally, when agents operate according to 
a grassroots protocol (\ia) the possible behaviors and interactions of an isolated community of agents are not constrained by placing the community within a larger community and (\ib)  agents in an isolated community, when placed within a larger community, may have behaviors and interactions not possible when in isolation. In particular, agents may interact with each other across community boundaries. This notion supports the grassroots deployment of a distributed system -- autonomous independent instances deployed at different locations and over time that can possibly (but not necessarily) interoperate once interconnected.

\mypara{Asynchronous distributed multiagent transition systems} We formalize this intuitive notion of a grassroots protocol using asynchronous
distributed multiagent transition system~\cite{shapiro2021multiagent}.    Multiagent transition systems allow the description of faults, including safety (Byzantine) and liveness (fail-stop/network/Byzantine) faults and fault-resilient implementations,
and allow proving the resilience of specific protocols to specific faults, e.g. having less than one third of the agents be Byzantine and having the adversary control network delay in asynchrony.  In addition, any implementation among transition systems can be proven to be resilient to specific faults.  Thus, while the definition of a grassroots implementation below does not specifically address fault resilience, any implementation, including grassroots ones, can be proven resilient to specific faults
See Appendix \ref{section:preliminaries} for a brief introduction; we repeat below essential definitions and results.

Assume a set $\Pi$ of agents, each equipped with a self-assigned single and unique key-pair, and identify an agent  $p \in \Pi$ by its public key.  Sybils attacks, namely multiple agents operated by the same  underlying entity~\cite{shahaf2020genuine}, are discussed below.
While the set of all agents $\Pi$ could in principle be infinite (think of all the agents that are yet to be born), when we refer to a particular set of agents $P \subseteq \Pi$ we assume $P$ to be finite. We use $\subset$ to denote the strict subset relation.

\begin{definition}[Local States, $\prec$, Initial State]
A \temph{local states function} $S$ maps every set of agents $P \subseteq \Pi$ to a set of local states $S(P)$, satisfying
$S(\emptyset)=\{s0\}$, with $s0$ referred to as the \temph{initial local state}, and 
$P \subset P' \implies S(P) \subset S(P')$. The local states function $S$ also has an associated partial order $\prec_{S}$ on $S(\Pi)$,  abbreviated $\prec$  when $S$ is clear from the context,
with $s0$ as its unique  least element.
\end{definition}

Note that a member of $S(P)$ is a possible local state of an agent in $P$. Intuitively, think of $S(P)$ as the set of all possible sequences of messages among members of $P$, or the set of all possible  blockchains created and signed by members of $P$,  both with $\prec_S$ being the prefix relation and $s0$ being the empty sequence; or the set of all possible sets of posts/tweets and threads of responses to them by members of $P$, with $\prec_S$ being the subset relation, and $s0$ being the empty set.

% \andy{ Let me try a version of the above.} 

% \andy{\textbf{Definition}. We consider a fixed value $Z$, which is the infinite set of all possible \temph{local states}. A \temph{local states function} $S$ maps every set of agents $P \subseteq \Pi$ to a set of local states $S(P)\subseteq Z$. A local states function $S$ also has an associated partial order $\prec_{S}$ on $Z$ with no maximal elements, and with a unique least element $s0(S)$, abbreviated $s0$ when $S$ is clear from the context, and referred to as the \temph{initial local state} of $S$. Each set of local states $S(P)$ includes $s_0$ and is unbounded in the inherited ordering. }

% \andy{Intuitively, one can think of $Z$ as being the set of all possible sequences recording which messages are sent and recieved by an individual agent, with such sequences ordered in the standard way by the prefix relation. }

% \andy{Is there some extra condition required here to ensure $S(P)\neq Z$? Intuitively, $S(P)$ should only include things produced by members of $P$, but that doesn't seem to be required yet.}

In the following, we assume a given fixed local states function $S$ with its associated ordering $\preceq_S$ and initial state $s0$.
\begin{definition}[Configuration, Transition, $p$-Transition, $\prec$]\label{definition:transition}
Given a set $X \subseteq S(\Pi)$ that includes $s0$ and a finite set of agents $P\subseteq \Pi$, a \temph{configuration} $c$ over $P$ and $X$ is a member of $C:=X^P$, namely $c$ consists of a set of local states in $X$ indexed by $P$.  
Given a configuration $c \in C$ and an agent $p \in P$, $c_p$ denotes the element of $c$ indexed by $p$, referred to as the \temph{local state of $p$ in $c$},
and $c0:=\{s0\}^P$ denotes the \temph{initial configuration}.
A \emph{transition} is an ordered pair of configurations, written $c \rightarrow c' \in C^2$.  If $c_p \ne c'_p$ for some $p \in P$ and $c'_q = c_q$ for every $q \ne p \in P$, the transition is a \temph{$p$-transition}.
%Let $CC(P)$  denote the set of transitions over $P$.
%
A partial order  $\prec$ on local states induces a partial order on configurations, defined by $c \preceq c'$ if $c_p \preceq c'_p$ for every $p \in P$.  
\end{definition}
Normally, configurations over $P$ will have local states $X=S(P)$, but to define the notion of a grassroots system we will also use $X=S(P')$ for some $P' \supset P$.

\begin{definition}[Distributed Multiagent Transition System; Computation; Run]\label{definition:DMTS}
Given a set $X \subseteq S(\Pi)$ that includes $s0$ and a finite set of agents $P\subseteq \Pi$, a \temph{distributed multiagent transition system} over $P$ and $X$,
 $TS =(P,X,T,\lambda)$, has \temph{configurations} $C=X^P$; \temph{initial configuration}  $c0:=\{s0\}^P$;  a set of \temph{correct transitions} $T = \bigcup_{p \in P} T_p \subseteq C^2$, where each $T_p$ is a set of \temph{correct $p$-transitions}; and a \temph{liveness requirement} $\lambda \subset 2^T$ being a collection of sets of transitions.
A \temph{computation} of $TS$ is a potentially-infinite sequence of configurations over $P$ and $S$,  $r= c \xrightarrow{} c' \xrightarrow{}  \cdots $, with two consecutive configurations in $r$ referred to as a \temph{transition of $r$}.   A \temph{run} of $TS$ is a computation that starts with $c0$. 
\end{definition}

Multiagent transition systems are designed to address fault resilience, and hence need to specify faults, which are of two types -- safety (incorrect transitions) and liveness (infinite runs that violate a liveness requirement).  Def. \ref{definition:DMTS} summarizes what multiagent transition systems are, and Def. \ref{definition:ts-slc} below specifies safe, live and correct (safe+live) runs.  It also specifies the notion of a correct agent relative to a run.  All that an incorrect agent may do is violate safety or liveness (or both) during a run. 
Note that computations and runs may include  incorrect transitions.  The requirement that agent $p$ is live is captured by including $T_p$ in $\lambda$, and that all agents are live by  $\lambda = \{T_p : p \in P\}$.

\begin{definition}[Safe, Live and Correct Run]\label{definition:ts-slc}
Given a transition system  $TS=(P,X,T,\lambda)$, a computation $r$ of $TS$ is \temph{safe}, also $r \subseteq T$, if every transition of $r$ is correct.  We use  $c \xrightarrow{*} c' \subseteq T$ to denote the existence of a safe computation (empty if $c=c'$) from $c$ to $c'$. 
A transition $c'\rightarrow c''$ is \temph{enabled on $c$} if $c=c'$.
A run is \temph{live wrt $L \in \lambda$} if either $r$ has a nonempty suffix in which no transition in $L$ is enabled in a configuration in that suffix, or every suffix of $r$ includes a transition in $L$. A run $r$ is \temph{live} if it is live wrt every $L \in \lambda$.
A run $r$ is \temph{correct} if it is safe and live.
An agent $p \in P$ is \temph{safe in $r$} if $r$ includes only correct $p$-transitions;  is \temph{live in $r$} if for every $L \in \lambda$, s.t. $L \subseteq T_p$, $r$ is live wrt $L$; and  is \temph{correct} in $r$ if $p$ is safe and live in $r$. 
\end{definition}

Note that run is live if for every $L \in \lambda$ that is enabled infinitely often, the run has infinitely many $L$-transitions.  Also note that a finite run is live if for no $L\in \lambda$ there is an $L$-transition enabled on its last state.

A transition system is \emph{asynchronous} if progress by other agents cannot prevent an agent from taking an already-enabled transition.

\begin{definition}[Monotonic and Asynchronous Distributed Multiagent Transition System]\label{definition:multiagent-sa}
%\begin{definition}[Multiagent Distributed Transition Systems]\label{definition:multiagent}
A distributed multiagent transition system $TS=(P,X,T,\lambda)$ is 
\temph{monotonic} wrt a partial order $\prec$ if $c\rightarrow c' \in T$ 
implies that $c \prec c'$, and it is \temph{monotonic} if it is monotonic wrt $\prec_S$.  It is \temph{asynchronous} if it is monotonic and for every $p$-transition $c \xrightarrow{} c' \in T_p$, $T_p$ also includes every $p$-transition $d \xrightarrow{} d'$ for which  $c \prec_{S} d$ and
    $(c_p \rightarrow c'_p) = (d_p \rightarrow d'_p)$.
\end{definition}

\mypara{Grassroots protocol}  
Next, we define the notion of a protocol and a grassroots protocol.
\begin{definition}[Protocol]\label{definition:family}
A \temph{protocol} $\calF$ is a family of distributed multiagent transition systems over a local states function $S$ that has exactly one such transition system $TS(P) = (P,S(P),T(P),\lambda(P)) \in \calF$ over $P$ and $S$ for every
 $\emptyset \subset P \subseteq \Pi$.
%
%The \temph{states} of $\calF$ are $S(\calF) := \bigcup_{P \subseteq \Pi}S(P)$.The protocol $\calF$ is \temph{asynchronous} if every transition system in $\calF$ is asynchronous, and is a \temph{subset} of protocol $\calF'$ if for every $P\subseteq \Pi$, $TS(P)$ is a subset of $TS'(P)$.
\end{definition}

Note that in multiagent transition systems, safety and liveness are properties of runs, not of protocols.  Hence, when a protocol is specified with a multiagent transition system, there is no need to specify additional ``protocol specific'' safety or liveness requirements, as any such property is implied by the definition itself.

For simplicity and to avoid notational clutter, we often assume a given set of agents $P$, refer to the  member of $\calF$ over $P$ as a representative of the family $\calF$, and refer to the protocol $TS(P) = (P,S(P),T(P),\lambda(P)) \in \calF$ simply as $TS = (P,S,T,\lambda)$.
%

%We note that all example protocols presented in~\cite{shapiro2021multiagent} are not grassroots, and so are all client-server and cloud-based systems, as well as peer-to-peer systems that require all-to-all dissemination such as mainstream cryptocurrencies, standard consensus protocols, and pub/sub systems. Among peer-to-peer file systems, to the best of our knowledge BitTorrent with private trackers is grassroots, as different autonomous instances of the protocol may use independent private trackers that  may later agree to interoperate, whereas IPFS is not, as it is designed to be a single global network. We elaborate on this below.

\begin{definition}[Projection]\label{definition:projection}
Let $\calF$ be a protocol over $S$, $\emptyset \subset P \subset P' \subseteq \Pi$.  Given a configuration $c'$  over $P'$ and $S(P)$, the \temph{projection of  $c'$ on} $P$, $c'/P$, is the configuration $c$ over $P$ and $S(P)$ satisfying $c_p = c'_p$ for all $p \in P$. The \temph{projection of $TS(P') = (P',S(P'),T',\lambda') \in \calF$ on $P$}, denoted $TS(P')/P$ is the transition system over $P$ and $S(P')$, $TS(P')/P:=(P,S(P'),T'/P,\lambda'/P)$,
where  $c_1/P \rightarrow c_2/P \in T'/P$ if  $c_1 \rightarrow c_2 \in T'$ and with $\lambda'/P := \{L/P : L \in \lambda'\}$.  
\end{definition}
Note that $TS(P')/P$ has local states  $S(P')$, not $S(P)$.  This is necessary as, for example, if the local state is a set of blocks, and in a $TS(P')$ configuration $c$ the local states of members of $P$ have blocks produced by members of $P'\setminus P$, this remains so also in $c/P$.

\begin{definition}[Subset]
Given  transition systems $TS=(P,X,T,\lambda)$,  $TS'=(P,X',T',\lambda')$, then $TS$ is a \temph{subset} of $TS'$, $TS \subseteq TS'$, if  $X\subseteq X'$, $T \subseteq T'$, and $\lambda$ is $\lambda'$ restricted to $T$,  $\lambda := \{ L \cap T |  L\in \lambda'\}$.
\end{definition}

\begin{definition}[Grassroots]\label{definition:grassroots}
A  protocol $\calF$ is \temph{grassroots} if $\emptyset \subset P \subset P' \subseteq \Pi \text{ implies that } TS(P) \subset TS(P')/P$. 
\end{definition}

%\andy{These seems quite a ``fundamental'' definition, but it would be nice to see further justification for the idea that grassroots protocols are the ones we want to consider. For example, could we have a theorem that asserts something like ``If a protocol doesn't require knowledge of the player set and achieves some minimal functionality, then it must be ``grassroots''?} 

Namely, in a grassroots protocol, a group of agents $P$, if embedded within a larger group $P'$, can still behave as if it is on its own (hence the subset relation), but also has new behaviors at its disposal (hence the subset relation is strict), presumably due to interactions between members of $P$ and members of $P'\setminus P$.

Appendix \ref{appendix-section:protocol-AD} proves that the All-to-All Dissemination protocol $\calA\calD$~\cite{shapiro2021multiagent} is not grassroots. The formal proof can be generalized, informally, to any protocol that employs all-to-all dissemination:
\begin{observation}\label{observation:ata-not-grassroots}
An all-to-all dissemination protocol cannot be grassroots.
\end{observation}
\begin{proof}[Proof of Observation \ref{observation:ata-not-grassroots}]
Informally,  liveness of an all-to-all dissemination protocol requires any object (message, post, block) produced by one correct agent to be eventually received by all correct agents participating in the protocol. Hence, any  infinite correct behavior of agents $P$ executing  an all-to-all dissemination protocol on their own, when embedded within a larger group $P'$ that produces additional objects, violates liveness. The reason is that in this infinite behavior members of $P$ never receive objects created by $P'\setminus P$, which they should if run within the larger group $P'$.  Hence
$TS(P) \subseteq TS(P')/P$ does not hold, and neither the stricter $TS(P) \subset TS(P')/P$, and thus the protocol is not grassroots.
\end{proof}

\mypara{Sufficient condition for a protocol to be grassroots} 
Here we define certain properties of a protocol and prove that satisfying them is sufficient for the protocol to be grassroots.
\begin{definition}[Monotonic and Asynchronous Protocol]\label{definition:monotonic-protocol}
Let $\calF$ be a protocol over $S$.
Then  $\prec_S$ is \temph{preserved under projection} if for every $\emptyset \subset P \subset P' \subseteq \Pi$ and every two configurations $c_1, c_2$ over $P'$ and $S$, 
$c_1 \preceq_S c_2$ implies that $c_1/P \preceq_S c_2/P$.
The protocol $\calF$ is \temph{monotonic} if $\prec_S$ is preserved under projection and every member of $\calF$ is monotonic; it is \temph{asynchronous} if, in addition, every member of $\calF$ is asynchronous.
\end{definition}

A protocol is interactive if a set of agents embedded in a larger group of agents  can perform  computations they cannot perform in isolation.  It is non-interfering if (\ia) \emph{safety:} a transition that can be carried out by a group of agents can still be carried out if there are additional agents that are in their initial state, and (\ib) \emph{liveness:} if an agent is live in a run of the group, then adding more agents to the group and extending the run with their transitions, will not result in the agent violating liveness. Formally:

\begin{definition}[Interactive \& Non-Interfering Protocol]\label{definition:non-interfering}
A protocol  $\calF$ over $S$ is \temph{interactive} if
for every $\emptyset \subset P \subset P' \subseteq \Pi$, 
$TS(P')/P \not\subseteq TS(P)$.
It is \temph{non-interfering} if for every $\emptyset \subset P \subset P' \subseteq \Pi$, with transition systems $TS = (P,S,T,\lambda), TS' = (P',S,T',\lambda') \in \calF$:
\begin{enumerate}
    \item \textbf{Safety}: For  every transition $c1 \rightarrow c2 \in T$,
$T'$ includes the transition  $c1' \rightarrow c2'$ for which $c1=c1'/P$, $c2=c2'/P$, and
$c1'_p = c2'_p = c0'_p$ for every $p \in P' \setminus P$, and
    \item \textbf{Liveness}: For every agent $p \in P$ and run $r$ of $TS$, if $p$ is live in $r$ then it is also live in every run $r'$ of $TS'$ for which $r'/P = r$.
\end{enumerate}
\end{definition}

%\begin{definition}[Closure Under Union]
%Let $S$ be a set of local states over $\calA$ closed under set union, $P \subset \Pi$ and $C$ a set of configurations over $P$ and $S$.  Then the \temph{global state} of $C$ is defined to be $S(C) := \bigcup_{p \in P} c_p \in S$.
%\end{definition}

%Let $\calA$ be a base set and $S$ be a set of sets over $\calA$.  Then $S$  is \temph{closed under set union} if $s, s' \in S$ implies that $s \cup s' \in S$.  Note that Blockchains are not closed under set union.

\begin{restatable}[Grassroots Protocol]{theorem}{GrassrootsProtocol}\label{theorem:grassroots}
An asynchronous, interactive, and non-interfering protocol is grassroots.
\end{restatable}

We note that  blockchain consensus protocols with hardcoded miners, e.g. the seed miners of Bitcoin~\cite{bitcoin-p2p} or the bootnodes of Ethereum~\cite{ethereum-bootnodes},  as well as permissioned consensus protocols with a predetermined set of participants such as Byzantine Atomic Broadcast~\cite{shostak1982byzantine,keidar2023cordial}, are all interfering, as additional participants cannot be ignored.    If hardcoded bootnodes/seed  miners are relativized, two instances of the protocol would interfere on the assigned global port numbers, preventing a server from participating in two instances. If both bootnodes and port numbers are relativized, the resulting protocol would not be interactive, as two instances of the protocol with disjoint bootnodes and port numbers, say bitcoin$'$ and bitcoin$''$,  would not be able to interact.
Thus, both the original definitions of Ethereum and Bitcoin, as well as definitions relativizing bootnodes and port numbers, are not grassroots.   

However, Theorem \ref{theorem:grassroots} and the examples above do not imply that ordering consensus protocols cannot be grassroots.  The problem is not with the consensus protocol as such, but with the choice of its participants. If participants in an instance of an ordering consensus protocol are  determined by the agents themselves via a grassroots protocol, rather than being provided externally and \emph{a priori} as in standard permissioned consensus protocols, 
or are required to include a predetermined set of initial participants as in Bitcoin and Ethereum, then the participants can reach consensus without violating grassroots.  Devising grassroots consensus protocols is a subject of future work (see ~\cite{poupko2023thesis}, Ch. 4).

Next, we present the \GD protocol and prove it to be grassroots.

\subsection{The \GD Protocol}

The \AD protocol is asynchronous and interactive, and this is also true of \GD. But, unlike \AD, in  the \GD  protocol dissemination occurs only among friends, defined below.  Moreover, agents are free to choose their friends.  Hence, additional agents may be ignored by a given group of agents without violating agent liveness, implying that \GD is also non-interfering and thus grassroots, as proven below.

We assume a given \emph{payloads function} $\calX$ that maps each set of agents $P$ to a set of payloads $\calX(P)$.  For example, $\calX$ could map $P$ to all strings signed by members of $P$; or to all messages sent among members of $P$, signed by the sender and encrypted by the public key of the recipient; or to all financial transactions among members of $P$. Remember that here $P$ are not `miners' serving transactions by other agents, but are the full set of agents participating the in protocol.
 
\begin{definition}[Simple Block, $SB$, $\prec_{SB}$]
A \temph{block Id} over $P$ is a pair $\Id = (p,i)$ where $p\in P$ and $i \in \NN$.
A \temph{simple block} over $P$ is a pair $(\Id,x)$ where $\Id$ is a block Id over $P$ and 
$x \in \calX(P)$ or $x=(\textit{follow},q)$, $q\in P$.  Such a block with $\Id = (p,i)$ is referred to as an \temph{$i$-indexed simple $p$-block with payload $x$}.  Let $B(P)$ denote the set of all simple blocks over $P$.

The local states function $SB$ maps $P$ to the set of all sets of simple blocks over $P$.
The partial order $\prec_{SB}$ over configurations over $P$ and $SB$ is defined  by $c \preceq_{SB} c'$ if
   $c, c'$ are configurations over $P$ and $SB$ and $c_p \subseteq c'_p$ for every $p \in P$.
\end{definition}
Note that  $c\prec_{SB} c'$ if $c_p \subset c'_p$ for some $p \in P$ and $c_q = c'_q$ for every $q \ne p \in P$.

We say that agent $p$ \emph{knows} a block $b$ if $b$ is in $p$'s local state. The liveness condition of all-to-all dissemination ensures that every block created by every correct agent will be known eventually by every other correct agent. This is  unnecessary and impractical for grassroots applications such as grassroots social networking~\cite{shapiro2023gsn} and grassroots cryptocurrencies\cite{shapiro2024gc}, discussed in Section \ref{section:applications}, especially as they grow and become very large scale.  Therefore the \GD  protocol satisfies a weaker, more local liveness requirement. 

In \GD, agents may follow other agents, and two agents are friends if they follow each other.
Informally, the basic rule of \GD is that an agent $p$ can receive from agent $q$ a $q'$-block $b$ 
if $p$ and $q$ are friends and either $q=q'$ or they both follow $q'$.
Note  that the friendship relation induces an undirected graph on the agents.  A \emph{friendship path}  is a path in the that graph.  The key liveness claim of \GD is that  $p$ eventually knows any $q$-block if there is a friendship path of correct agents from $p$ to $q$, all of which follow $q$.

\begin{definition}[Follows, Needs, Friend, Social Graph]\label{defininition:follows}
Given $p,q \in P$ and a configuration $c$ over $P$ and $SB$:
$p$ \temph{follows} $q$  in $c$ if  $c_p$ includes a $p$-block with payload
 $(\textit{follow},q)$;
$p$ \temph{needs} the $q$-block $b$ in $c$ if $p$ follows 
$q$ in $c$ and $b\in c_q\setminus c_p$;
and $p$ and $q$ are \temph{friends} in $c$ if $p$ and $q$ follow each other in $c$.
The \temph{social graph} $(P,E(c))$ induced by $c$ has an edge $(p,q)\in E(c)$ if $p$ and $q$ are friends in $c$.
\end{definition}

The key idea of the Grassroots Dissemination protocol is:
\begin{tcolorbox}[colback=gray!5!white,colframe=black!75!black]
\textbf{Principle of \GD:} \\
An agent that needs a block can obtain it from a friend that has it. 
\end{tcolorbox}

\begin{definition}[$\calG\calD$: \GD]\label{definition:GD}
The \temph{\GD protocol} $\calG\calD$ over $SB$ has for each $P\subseteq \Pi$ the transition system $GD = (P,SB,T,\lambda)$, transitions $T$ with a $p$-transition $c \rightarrow c' \in T_p$ for every   $p \in P$, $c'_p = c_p \cup \{b \}$, $b= (\Id,x) \notin c_p$, and either:
\begin{enumerate}[partopsep=0pt,topsep=0pt,parsep=0pt]
    \item \textbf{Create}:  $\Id=(p,i)$, $i = \text{ max } \{ j \ge 0 : (p,j,x) \in c_p\}+1$, $x\in \calX(P)$, or 
    \item \textbf{$q$-Disseminates-$b$}:  $b \in c_q$ for some $q \in P$,  $p$ and $q$ are friends in $c$, and $p$ needs $b$ in $c$. 
\end{enumerate}
The liveness condition $\lambda$ includes for each $p,q \in P$ and $b \in B(P)$ the set of all $p$-transitions labeled $q$-Disseminates-$b$.
\end{definition}

Every agent $p$ can either add a consecutively-indexed $p$-block to its local state via a Create transition or obtain from a friend a $q$-block $b$ it needs via a Disseminate transition. A Create transition can also initiate following another agent $q$:  Since \textsc{follow} is just a `reserved word' with a specific interpretation by the protocol and $\calX(P)$ contains all strings, it also contains the string representation of $(\textsc{follow},q)$, for any agent $q$. The liveness condition ensures that every correct agent $p$ receives any block by an agent it follows, if the block is known to any of $p$'s friends.  As there are no other liveness requirements,  agents are free to choose which agents to follow, if at all.

\begin{restatable}[$\calG\calD$ Grassroots Liveness]{proposition}{GrassrootsLiveness}\label{proposition:GD-local-liveness}
Let $r$ be a run of GD, $p,q \in P$.
If there is a configuration $c\in r$ such that $p$ and $q$ are connected via a friendship path, 
$p_1,p_2,\ldots,p_k$, $q=p_1$, $p=p_k$, $k\ge 1$ all of its members are correct in $r$ and  follow $q$ in $c$ then for every  $q$-block $b$ in $c$  there is a subsequent configuration $c' \in r$ for which 
$b \in c'_p$.
\end{restatable}

%The fault-resilience of the composition of fault-resilient implementations has yet to be formalized and investigated.  If the bottom protocol of a protocol stack is resilient to certain faults, it can be concluded that the entire stack is resilient to such fault.  However, if may be the case that some faults are better described, detected and excluded by higher-level protocols.

\begin{restatable}[]{proposition}{GDgrassroots}\label{proposition:GD-grassroots}
$\calG\calD$ is grassroots.    
\end{restatable}

%We opted here for a simple notion of following an agent.  It can be  refined so that an agent be followed starting from a specific block and later be unfollowed, which may provide for a more efficient implementation of grassroots currencies~\cite{shapiro2024gc,lewis2023grassroots}.

\subsection{Grassroots Implementation}
Here we define the notion of a grassroots implementation and prove that having a grassroots implementation is a sufficient condition for a protocol to be grassroots.
First we recall the notion of implementation among multiagent transition systems~\cite{shapiro2021multiagent} (See also Appendix \ref{section:preliminaries}). 

\begin{definition}[Specification; Implementation: Piecemeal, Safe, Live, Correct and Complete]\label{definition:implementation}
Given two distributed multiagent transition systems, $TS=(P,S,T,\lambda)$ over $P$ and $S$ (the \temph{specification}) and $TS'=(P,S',T',\lambda')$ over $P$ and $S'$, \temph{an implementation of $TS$ by $TS'$} is a function $\sigma : C' \rightarrow C$ where $\sigma(c0') = c0$, in which case the pair $(TS', \sigma)$ is referred to as  \temph{an implementation of $TS$}. The implementation is \temph{piecemeal} if  $\sigma$ is defined for local states and then
extended to configurations by  $\sigma(c')_p := \sigma(c'_p)$ for $c'\in C'$ and $p \in P$.
Given a computation $r'= c'_1\rightarrow c'_2 \rightarrow \ldots$ of $TS'$, $\sigma(r')$ is the (possibly empty) computation obtained from the computation $\sigma(c'_1)\rightarrow \sigma(c'_2) \rightarrow \ldots$ by removing consecutively repetitive elements so that $\sigma(r')$ has no \temph{stutter transitions} of the form $c \rightarrow c$. The implementation $(TS', \sigma)$ of $TS$ is \temph{safe/live/correct} if $\sigma$ maps every safe/live/correct $TS'$ run $r'$ to a safe/live/correct $TS$ run $\sigma(r')$, respectively, 
and is \temph{complete} if every correct run $r$ of $TS$ has a correct run $r'$ of $TS'$ such that $\sigma(r')=r$.
\end{definition}
Note that while an implementation in general is defined for configurations, a piecemeal implementation is defined for local states, which entails its definition for configurations.   Not all implementations are piecemeal, but, as shown next, piecemeal and grassroots  go hand-in-hand.

The standard way of proving that an implementation of one multiagent transition system (the specification) by another is fault resilient is by showing that the implementation maps any incorrect run of the implementing transition system (with specific types of safety and liveness faults by incorrect agents; a typical example would be runs in which the number of faulty agents is less than one third of all agents) to correct runs of the specification. 
\begin{definition}[Grassroots Implementation]\label{definition:grassroots-implementation}
Let $\calF, \calF'$ be protocols over $S, S'$, respectively,  and $\sigma$ a correct and complete implementation of $TS'(P)$ by $TS(P)$ for every $P \subseteq \Pi$.  Then $(\calF, \sigma)$ is an implementation of $\calF'$.  If, in addition, $\calF$ is grassroots and $\sigma$ is piecemeal, then $(\calF, \sigma)$ is a \emph{grassroots implementation} of $\calF'$.
\end{definition}

The following theorem shatters the hope of a non-grassroots protocol to have a grassroots implementation, and can be used to prove that a protocol is grassroots by providing it with a grassroots implementation.
\begin{theorem}[Grassroots Implementation]\label{theorem:grassroots-implementation}
A protocol that has a grassroots implementation is grassroots.
\end{theorem}
\begin{figure}[ht]
\centering
\includegraphics[width=13cm]{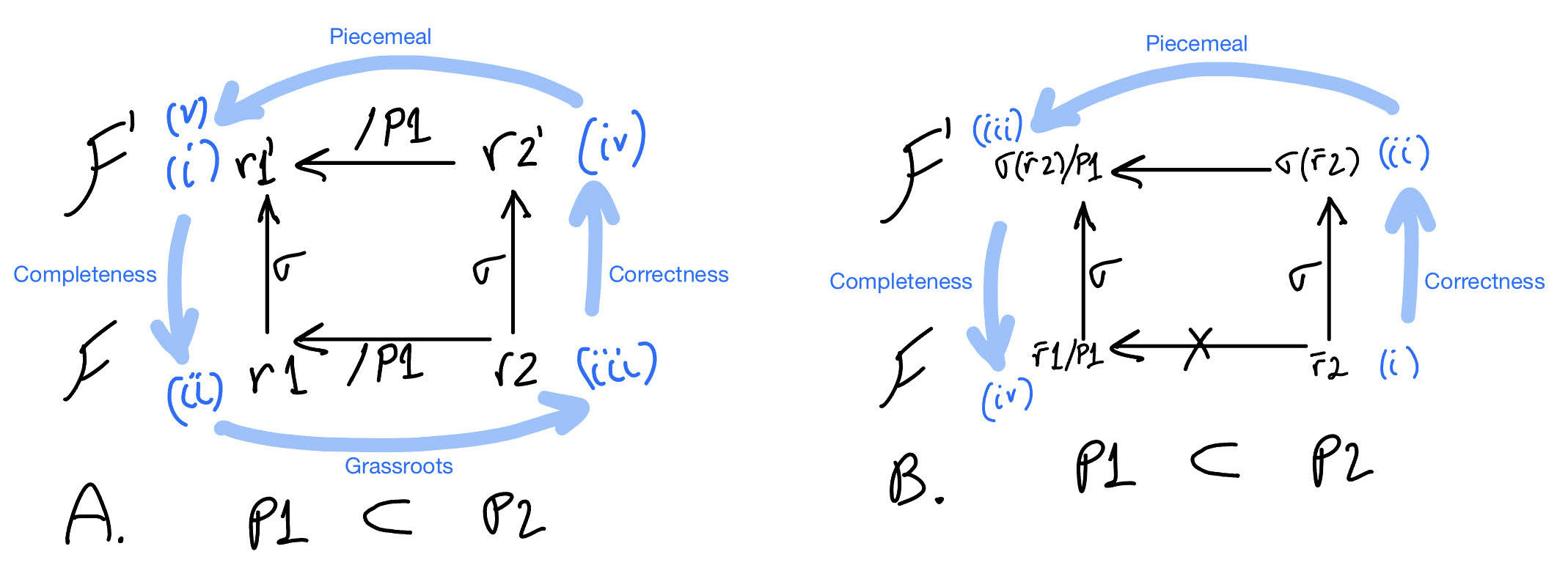}
\caption{Some Steps in the Proof of Theorem \ref{theorem:grassroots-implementation}. A. Proof of $TS'(P1)\subseteq TS'(P2)/P1$  B.  Proof of $TS'(P1)\not\supseteq TS'(P2)/P1$ }
\label{figure:grassroots-implementation}
\end{figure}

\begin{proof}[Proof of Theorem \ref{theorem:grassroots-implementation}]
Let $(\calF,\sigma)$ be a grassroots implementation of protocol $\calF'$.  We wish to prove that $\calF'$ is grassroots, namely that for $\emptyset \subset P1 \subset P2 \subseteq \Pi$,  $TS'(P1) \subset TS'(P2)/P1$.   First we prove that  $TS'(P1) \subseteq TS'(P2)/P1$. 
Roman numerals refer to Figure \ref{figure:grassroots-implementation}.A.
Consider (\ia) a correct run $r1'$ of $TS'(P1)$.  Since $\sigma$ is complete, there is (\ib) a correct run $r1$ of $TS(P1)$ such that $r1' = \sigma(r1)$.  Since $\calF'$ is grassroots, there is (\ic) a correct run $r2$ of $TS(P2)$ such that $r2/P1 = r1$.  Since $\sigma$ is a correct implementation of $\calF'$ by $\calF$, then (\id) $r2' = \sigma(r2)$ is a correct run of $TS$.
Since $\sigma$ is piecemeal by assumption, for every configuration $c \in r2$ and every $p \in P1$, $\sigma(c)_p =\sigma(c_p)$ and also $\sigma(c/P)_p = \sigma((c/P)_p) = \sigma(c_p)$.  Hence (\iie) $r2'/P1 = \sigma(r2)/P1 = \sigma(r2/P1) = \sigma(r1) = r1'$.  Namely $r1'$ is a run of $TS'(P2)/P1$, concluding that $TS'(P1)\subseteq TS'(P2)/P1$.

Next we prove that $TS'(P1) \not\supseteq TS'(P2)/P1$. 
Roman numerals refer to Figure \ref{figure:grassroots-implementation}.B.
Since $F$ is grassroots, there is (\ia) a run $\bar{r}2$ of $TS(P2)$ such that
$\bar{r}2/P1$ is not a run of  $TS(P1)$.  Since $\sigma$ is correct, then (\ib) $\sigma(\bar{r}2)$ is a run of $TS'(P2)$.  We prove (\ic) that $\sigma(\bar{r}2)/P1$ is not a run of $TS'(P1)$ by way of contradiction.  Assume it is.  
Since $\sigma$ is piecemeal, it follows that $\sigma(\bar{r}2)/P1= \sigma(\bar{r}2/P1)$.  Then by completeness of $\sigma$, it follows that 
$\bar{r}2/P1$ is a run of $TS(P1)$, a contradiction.
This completes the proof, concluding the $\calF'$ is grassroots.
%\qed
\end{proof}

\section{Grassroots Implementation of Grassroots Dissemination}\label{section:GD-implementation}

Here we present an implementation of the \GD protocol $\calG\calD$,
as depicted in Figure \ref{figure:GD-stack}:  The blocklace-based \CD  Protocol $\calCD$ (Def. \ref{definition:protocol-CD}),   and its proof of being grassroots (Prop.\ref{proposition:CD-grassroots});  the implementation $\sigma$ (Def.\ref{definition:sigma-CD-GD}) of $\calG\calD$ by $\calCD$, and its proof of being grassroots (Thm. \ref{theorem:CD-implements-GD}); and algorithm \ref{alg:blocklace} with pseudocode for blocklace utilities and Algorithm \ref{alg:CD} with pseudocode realizing $\calCD$ for the model of Asynchrony.

The \CD protocol employs the notions of \emph{follows}, \emph{needs}, \emph{friend}, and \emph{social graph} similarly to $\calG\calD$ (Def. \ref{defininition:follows}).
A key difference is that in \GD an agent that needs a block $b$ may receive it from a friend that has $b$ `by magic',  whereas in \CD such an agent $p$ must first disclose to a friend $q$ the blocks it knows and the agents it follows, implying that it needs $b$, and based on such disclosure $q$ may cordially send $b$ to $p$. 

%\andy{As I read through for the first time, it feels there is something slightly jarring going on. The \GD protocol already seems obviously grassroots, the thing that seems missing is an explanation of how the state transitions required for liveness can actually be realised by a detailed (real life) program: How do my friends know what to send me? But the way things are presented here, there doesn't yet seem to be any formal difference between the status of \GD and \CD (in terms of one being sufficiently detailed to actually count as a full specification). Instead, it seems that \CD is being used to prove that \GD is grassroots.}
%\udi{Indeed the specification is very abstract, and it can be proven grassroots in two ways - directly or via a grassroots implementation. Alan Perlis had a saying, that one programmer's constant is another programmer's variable.   All transition systems are abstract, some more than others.  Pseudocode is considered sufficiently concrete, and that's the bottom implementation, even if not proven formally.  Does this answer your comment?}

\begin{tcolorbox}[colback=gray!5!white,colframe=black!75!black]
\textbf{Principles of \CD:}
\begin{enumerate}[leftmargin=*]
    \item \textbf{Disclosure}: Tell your friends which blocks you know and which you need
    \item \textbf{Cordiality}: Send to your friends blocks you know and think they need
\end{enumerate}
\end{tcolorbox}
We employ the blocklace~\cite{keidar2021need} to realize these principles. The blocklace is a DAG-like partially-ordered generalization of the totally-ordered blockchain data structure that  functions as a fault-resilient, conflict-free replicated data type~\cite{shapiro2011conflict}.
While the mathematical construction and proofs of the blocklace realization presented next are quite involved, the final result, presented as 40 lines of pseudocode in Algs. \ref{alg:blocklace}, \ref{alg:CD} and \ref{alg:CD-UDP} is quite simple.

\subsection{Blocklace Preliminaries}

Each block in a blocklace may contain a finite set of cryptographic hash pointers to previous blocks, in contrast to one pointer (or zero for the initial block) in a blockchain.  We assume a payload function $\calX$ as above, and a cryptographic hash function \emph{hash}.

\begin{definition}[Block, $p$-Block, $p$-Pointer, Self-Pointer, Initial Block]\label{definition:block}
%, $p$-Chain]\label{definition:block}
A \temph{block} over $P \subseteq \Pi$ is a triple $b=(h^p,H,x)$, referred to as a \temph{$p$-block}, $p \in P$, with $h^p$, referred to as a \temph{$p$-pointer},  being the hash pointer $\textit{hash}((H,x))$ signed by $p$; $x \in\calX(P)$ being the \temph{payload} of $b$; and $H$ a finite set of signed hash pointers.   For each $h^q \in H$ there is a $q$-block $b'=(h^q,H',x')$ over $P$ and $\calX$, in which case we say that $b$ \temph{points to} $b'$, and if $p=q$ then $h^q \in H$ is a \temph{self-pointer}. 
If $H = \emptyset$ then $b$ is \temph{initial}. 
%
% A \temph{chain} (or \temph{personal blockchain}) is a sequence of blocks in which the first is initial and each of the subsequent blocks includes a self-pointer to its predecessor; it is a $p$-chain if it consists of $p$-blocks.
%
The set of blocks over $P$ is denoted $\calB(P)$. A \temph{blocklace} over $P$ is a subset of $\calB(P)$. The set of all blocklaces over $P$ is denoted by $\calS\calB(P)$.
\end{definition}
\mypara{Notes} 
(\ia) The definition of a cryptographic hash function implies that it infeasible for a computationally-bounded adversary to compute a set of blocks that form a cycle.
(\ib) A signed hash pointer informs of the creator of the pointed block and its authenticity without revealing the contents of the block.  Note that the blockchain and blocklace use hash pointers for exactly the same reason: So that the resulting data structure is tamper proof and agents cannot repudiate having produced a block.
(\ic) As a cryptographic hash function is collision-free WHP, 
in the following we use the $p$-pointer $h^p$ of a block $(h^p,H,x)$ as the block's name.
(\id) In the following protocols, the non-initial blocks created by a correct agent have exactly one self-pointer and form a single chain that ends in an initial block. 
    %\item In principle, different agents, even different blocks, may use different hash functions, provided pointers are extended so that which hash function a pointer uses is provided by the pointer.  This facilitates a more flexible grassroots deployment, as different protocol instances may employ different hash functions, as well as more flexible protocol upgrade, as agents may change the hash function they use midstream provided the change is disclosed.
%When it does not cause confusion we abbreviate $\calB(P)$ as $\calB$.

A blocklace  $B \subseteq \calB(P)$ induces a finite-degree directed graph $(B, E)$, $E \subset B \times B$, with blocks in $B$ as vertices and directed edges $(b,b') \in E$ if $b \in B$ includes a hash pointer to $b' \in B$. We overload $B$ to also mean its induced directed graph $(B,E)$, and following standard DAG terminology refer to a block with zero in-degree in a blocklace as a \temph{source} of the blocklace.  Note that the directed graph induced by any blocklace is acyclic. 

\begin{definition}[$\succ$, Observe]\label{definition:observe}
The partial order $\succ$ over $\calB(P)$ is defined by $b'\succ b$ if there is a nonempty path from $b'$ to $b$ in $\calB(P)$.  
A block  $b'$ \temph{observes} $b$ if $b'\succeq b$. Agent $p$ \temph{observes $b$ in} $B$ if there is a $p$-block $b' \in B$ that observes $b$. 
%, and a group of agents $Q \subseteq \Pi$ \temph{observes $b$ in} $B$ if every agent $p \in Q$ observes $b$.
%
%The \temph{closure} of a block $b$ is $[b] := \{b' : b\succeq b'\}$ and the \temph{closure} of a blocklace $B$ is $[B] := \bigcup_{b\in B} [b]$.  A blocklace $B$ is \emph{closed} if $B = [B]$.
\end{definition}
%This notion of closure is employed extensively by blocklace protocols that require all-to-all dissemination, e.g. the Cordial Miners consensus protocols~\cite{keidar2023cordial}. Grassroots dissemination requires a more refined notion of closure for the local blocklace of an agent, presented below.
%

\subsection{The \CD Protocol}
%$\calCD$}

\begin{wrapfigure}{r}{5cm}
  \begin{center}
   \includegraphics[width=5cm]{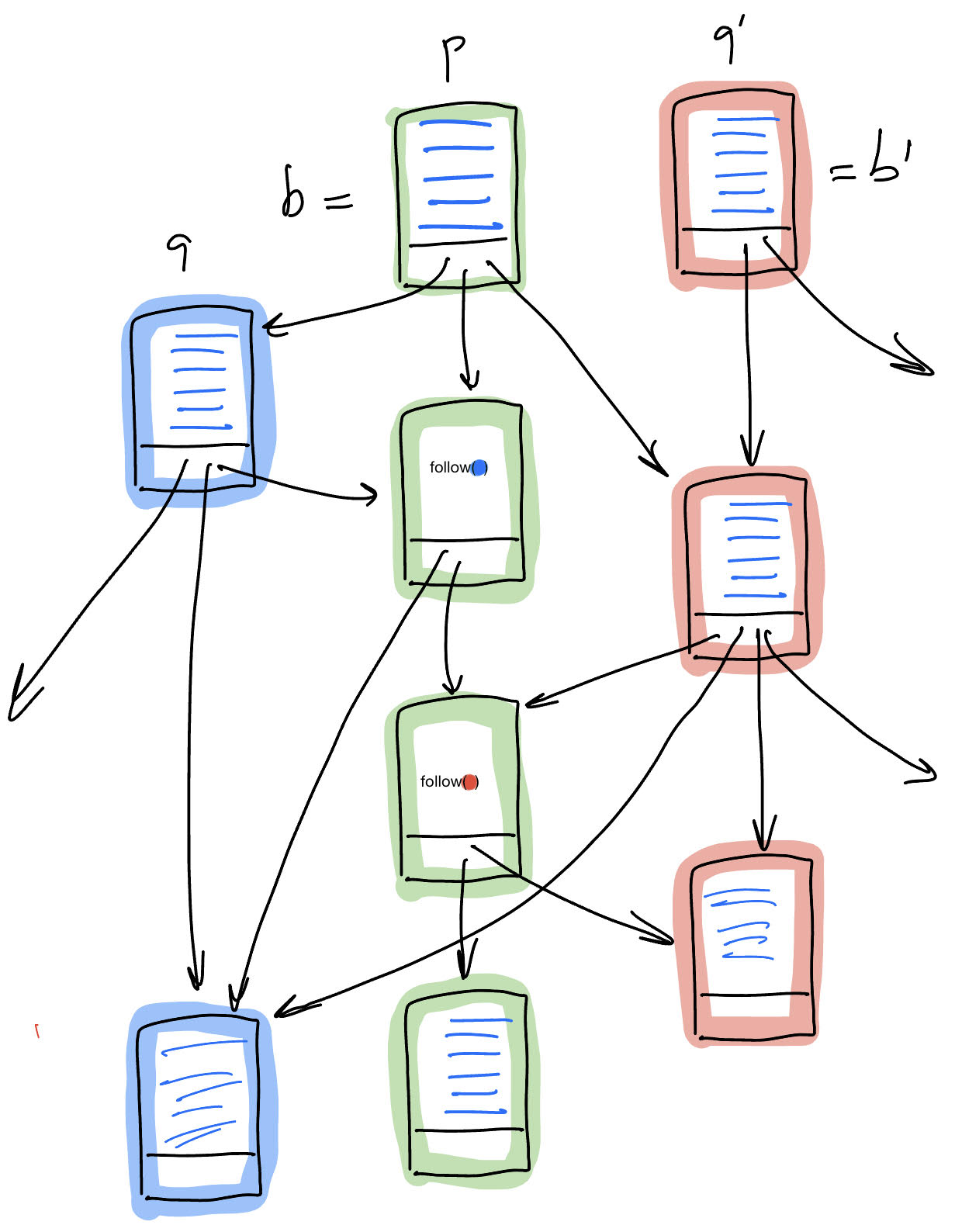}
  \end{center}
  \caption{$p$-Closure and $p$-Closed (Def. \ref{definition:p-closure}): Blocks are color-coded by agent, thus pointers among blocks of the same color are self-pointers, others are non-self pointers. Specific payload is shown when relevant.  The $p$-closure of the $p$-block $b$ includes the self-path from $b$, the $q$ self-path starting from the $q$-block  pointed to by $b$,  and the $q'$ self-path starting from the (second from top) $q'$-block  pointed to by $b$, all ending in their respective initial blocks.  Non-self pointers are left dangling in a $p$-closure if from $q$- and $q'$-blocks, but not from $p$-blocks.  
  Still, the entire blocklace shown in the figure, including the top $q'$-block, is $p$-closed.
  }
\label{figure:p-closure}
\end{wrapfigure}

The \CD protocol $\calCD$ realizes the two principles as follows.
Disclosure is realized by a new $p$-block serving as a \emph{multichannel ack/nack message}, informing its recipient whether it follows another agent $q$, and if so also of the latest $q$-block known to $p$, for every $q \in P$. 
This is realized by the notions defined next and illustrated in Figure \ref{figure:p-closure}.

\begin{definition}[Closure, Self-Closed, Closed, $p$-Closed, Maximally-Closed]\label{definition:p-closure}
Given a blocklace $B$ and a block $b \in \calB(P)$, the \temph{closure}  $[b]_B$ of $b$ in $B$ is the set of blocks each of which reachable from $b$ in $B$ via a (possibly-empty) path with at most one non-self pointer. We abbreviate $[b]_B$ as $[b]$ if $B$ is clear from the context. The block $b$ is \temph{self-closed} in $B$ if the maximal path of self-pointers from $b$ in $B$ ends in an initial block, and \temph{closed} in $B$ if every block in $[b]$ is self-closed. The blocklace $B$ is \temph{$p$-closed} if every $p$-block in $B$ is closed in $B$. 
A $p$-block closed in $B$ is \temph{maximally-closed} in $B$ if it observes every self-closed $q$-block in $B$ for every $q\in P$ followed by $p$ in $B$.
\end{definition}
In the following protocol agents need to know the blocks of their friends (reachable via one non-self pointer), but not the blocks of the friends of their friends (reachable only via two non-self pointers).  Hence the notion of $p$-closure defined above.
As we shall see, a correct agent $p$ always maintains a $p$-closed blocklace and produces only maximally-closed $p$-blocks, thus facilitating abiding by the principle of disclosure.

The local state in the protocol presented includes a blocklace and block messages.  

\begin{definition}[Blocklace Message, Blocklace Messaging Local States Function and Configuration, Blocklace of a Configuration]\label{definition:blocklace-configuration}
Given $P \subseteq \Pi$, a \temph{blocklace message} over $P$ is a pair $(p,b)$, with $p \in P$ being its \temph{destination} and $b\in \calB(P)$  its \temph{payload}. 
The \temph{blocklace messaging local states function} \textit{MSB}  maps $P$ to the set of all pairs $(B,Out)$, where $B$ is a finite subset of  $\calB(P)$ and $Out$ a finite set of blocklace messages over $P$.  Here,
$B$ is referred to as the \emph{local blocklace} and  $Out$ as the \emph{output messages} of the local state.  The associated partial order is defined by  $(B,Out) \preceq_{MSB} (B',Out')$
if $B \subseteq B'$ and $Out \subseteq Out')$.
A \temph{blocklace messaging configuration} $c$ over agents $P$ is a 
configuration over $P$ and \textit{MSB}.
The blocklace of a blocklace messaging configuration $c$ is defined by $B(c) := \bigcup_{p \in P, c_p=(B_p,Out_p)} B_p$.
\end{definition}

For a configuration $c$ over $P$ and \textit{MSB}, a block $b$ and $p \in P$ we use $b \in c_p$ for
$c_p=(B_p,Out_p)~ \&~ b \in B_p$.  With this abbreviation, definition of follows, needs, friend, and social graph for the blocklace messaging configurations is identical to their definition for simple blocks and its configurations (Def. \ref{defininition:follows}), except for replacing the simple blocks local state function $SB$ by the blocklace messaging local state function \textit{MSB}.

A key advantage of the blocklace is that agents may know global properties based on their local blocklace, provided these properties are monotonic wrt the subset relation (and thus hold globally, for the union of all local states, if they hold locally), as follows:
\begin{definition}[Agent Knowledge]\label{definition:agent-knowledge}
Given a blocklace configuration $c$ over $P$, an agent $p \in P$ \temph{knows a block $b$} in $c$ if $b \in c_p$, and $p$ \temph{knows a property} of the blocklace of configuration $c$ if the property is monotonic wrt the subset relation and holds in its local blocklace $c_p$.  In particular, $p$ \temph{knows that $q$ knows $b$} in $c$ if $c_p$ has a $q$-block that observes $b$; \temph{$p$ knows that $q$ follows $p'$ in $c$} if $c_p$ has a $q$-block that observes a $p'$-block;  and \temph{$p$ knows that $p'$ and $q$ are friends in $c$} if $p$ knows that $p'$ follows $q$ in $c$ and $p$ knows that $q$ follows $p'$ in $c$ (with $p=p'$ a special case).
\end{definition}

Pseudocode realizing basic blocklace utilities needed for grassroots dissemination are presented as Algorithm \ref{alg:blocklace}.

\begin{figure}[t]
\begin{algorithm}[H]
    \caption{\textbf{Cordial Dissemination Blocklace Utilities.} Code for agent $p $}
    \label{alg:blocklace}
    \small
    %\begin{changemargin}{0cm}{-2cm}
    \begin{algorithmic}[1] \scalefont{0.9}
        % \begin{multicols}{2}
        % \State \oded{I've decreased the font size for the algorithm to fit in.}
        \Statex \textbf{Local variables:}

        \StateX struct $\textit{h}$: \Comment{The structure of a hash pointer $h$ in a block, Def.\ \ref{definition:block}}
        \StateXX $h.\textit{hash}$ -- the value of the hash pointer
        \StateXX $h.\textit{creator}$ -- the signatory of the hash pointer
        
        \StateX struct $\textit{block } b$: \Comment{The structure of a block $b$ in a blocklace, Def.\ \ref{definition:block}}
        % \StateXX $b.\textit{round}$ - the round of $v$ in the DAG
        \StateXX $b.h$ -- hash value of the rest of the block, signed by its creator $p$
        \StateXX $b.\textit{payload}$ -- a set of transactions
  %      \StateXX $b.\textit{share}$ -- a share of the random coin, if needed 
        \StateXX $b.\textit{pointers}$ -- a possibly-empty set of signed hash pointers to other blocks

	  \StateX \textit{blocklace} $\gets \{\}$  \Comment{The local blocklace of agent $p$}
   
 		\vspace{0.5em}
		\Procedure{$\textit{create\_block}$}{\textit{payload}}   \Comment{1. \textbf{Create\&Disclose}}
		\State \textbf{new} $b$  \Comment{Allocate a new block structure}
		\State $b.\textit{pointers} \gets \{h :  h.\textit{hash} := b.\textit{hash} \wedge 
            h.\textit{creator} := b.\textit{creator} \wedge \textit{follows}(p,b.\textit{creator}) \wedge \textit{maximal(b)}\}$ \label{alg:sources} \Comment{Def. \ref{definition:p-closure}, Fig. \ref{figure:p-closure}}
          \State $b.\textit{payload} \gets \textit{payload}$ 
          \State $b.h \gets \textit{hash}((b.\textit{pointers},b.\textit{payload}))$ signed by $p$
	       \State $\textit{blocklace} \gets \textit{blocklace} \cup \{b\}$  \label{alg:BDA-create-next-block}
		\EndProcedure

 \vspace{0.5em}  
	    \Procedure{\textit{initial}}{$b$} \label{alg:friend}
       \Comment{$b$ is an initial block.  Def. \ref{definition:block}} 
            \State \Return 
            $b.\textit{pointers}=\emptyset$.
        \EndProcedure

        % \vspace{0.5em}
        \Procedure{$\textit{observes}(b,b')$}{} \Comment{Def.\ \ref{definition:observe}} 
        \State \Return $\exists b_1,b_2,\ldots,b_k \in \textit{blocklace}$, $k\ge 1$,  s.t.\
        $b_1 = b $, $b_k = b'$ and $\forall i \in [k-1] \exists h \in b_i.\textit{pointers} \colon h.\textit{hash}= \textit{hash}(b_{i+1})$
        \EndProcedure

        % \vspace{0.5em}
        \Procedure{$\textit{maximal}(b)$}{} \Comment{Def.\ \ref{definition:p-closure}} 
        \State \Return $\lnot\exists b'\in \textit{blocklace}: (b'\ne b \wedge
        b'.\textit{creator} = b.\textit{creator}  \wedge \textit{observes}(b',b))  \wedge$ 
        \StateX 
        $\exists b'\in \textit{blocklace}: (b'.\textit{creator}=b.\textit{creator} \wedge 
        \textit{initial}(b') \wedge \textit{observes}(b,b'))$
        \EndProcedure

        % \vspace{0.5em}
        \Procedure{$\textit{sources}$}{} \Comment{Def.\ \ref{definition:p-closure}} 
        \State $b^* \gets b\in \textit{blocklace}$ s.t. $b.\textit{creator} = p \wedge \textit{maximal}(b)$
        \State \Return $\{b^*\} \cup \{b \in blocklace \setminus [b^*] :
        \textit{maximal}(b)\}$
        \EndProcedure

        % \vspace{0.5em}

	 \Procedure{\textit{agentObserves}}{$q,b$} \label{alg:agentObserves}
       %\Comment{foo} 
            \State \Return $\exists b' \in \textit{blocklace} :
      b'.\textit{creator}=q \wedge \textit{observes}(b',b)$  \Comment{Def.\ \ref{definition:observe}, Fig. \ref{figure:p-closure}}
        \EndProcedure

\vspace{0.5em} 

	 \Procedure{\textit{follows}}{$q,q'$} \label{alg:follows}
       \Comment{Follows, Def.\ \ref{defininition:follows}, Fig. \ref{figure:p-closure}}
            \State \Return $ q= q' \vee \exists b \in \textit{blocklace}: 
      b.\textit{creator}=q 
     \wedge  b.\textit{payload}=(\textit{`follow'},q')$  
        \EndProcedure

\vspace{0.5em} 

	 \Procedure{\textit{friend}}{$q$} \label{alg:friend}
            \Return $\textit{follows}(p,q) \wedge \textit{follows}(q,p)$  \Comment{Def.\ \ref{defininition:follows}, Fig. \ref{figure:p-closure}}
        \EndProcedure

        \alglinenoNew{counter}
        \alglinenoPush{counter}
    % \end{multicols}
 
    \end{algorithmic}
     % \end{changemargin}
\end{algorithm}
\vspace{-2.5em}
\end{figure}

\begin{definition}[$\calCD$:  Cordial  Dissemination]\label{definition:protocol-CD}
The $\calCD$ \temph{cordial dissemination} protocol has for each $P \subseteq \Pi$  a distributed transition system $\text{CD}=(P,\textit{MSB},T,\lambda)$ with transitions $T$  being all $p$-transitions $c\xrightarrow{}c'$, where $p \in P$, $c_p= (B,Out)$ and one of the following holds:
\begin{enumerate}[partopsep=0pt,topsep=0pt,parsep=0pt]
    \item[1.]  \textbf{Create\&Disclose}:  $c'_p := (B \cup \{b\},Out)$ and $b \in \calB(P)\setminus B$ is a $p$-block maximally-closed in $B$, or
    \item[2.] \textbf{Send-$b$-to-$q$}: $c'_p :=(B,Out \cup \{(q,b)\})$, $b \in B$ 
    \item[3.] \textbf{Receive-$b$}:   $c'_p := (B \cup \{b\},Out)$, $b\notin B$, and for some $q \in P$, $(p,b) \in Out_q$,  $c_q=(B_q,Out_q)$.
    \item[4.] \textbf{Cordially-Send-$b$-to-$q$}: $c'_p :=(B,Out \cup \{(q,b)\})$,  $b \in B$,  $(q,b) \notin Out$, and (\ia) $p$ knows  that $q$ is a friend,  (\ib) $p$ knows that $q$ follows the creator of $b$,  (\ic) $p$ does not know that $q$ knows $b$.
\end{enumerate}
The liveness condition $\lambda$ partitions the $p$-transitions Create\&Disclose, Cordially-Send, and Receive according to their labels, so that all transitions with the same label are in the same partition.
\end{definition}
\mypara{Notes} Disclosure is realized by the Create\&Disclose transition choosing a maximally-closed $p$-block, and cordiality is realized by the Cordially-Send-$b$-to-$q$, with which $p$ sends any $p'$-block $b$ it knows to any friend $q$ that $p$ believes needs it. Namely, for which $p$ knows that $q$ follows $p'$ and $p$ doesn't know that $q$ knows $b$. If $q$ knows $b$ then $p$ would eventually know that, when it receives the next $q$-block that discloses knowledge of $b$.  The special case of Cordially-Send, where $p=p'$ and $b$ is a $p$-block, is simply dissemination of blocks created by $p$ to $p$'s friends. 

The liveness condition requires an agent to disclose every so often the blocks they know and receive any block sent to it that it does not know already, but it does not require an agent to follow other agents.
Two agents can send each other \textsc{follow} blocks to become friends. 
Unlike in $\calG\calD$, where $p$ can construct a simple initial $q$-block, here the initial $q$-block includes a hash pointer signed by $q$, which cannot be created/forged by $p$. 

\mypara{Asynchrony} The model of asynchrony assumes that each message sent among correct agents is eventually received and an that an adversary may control, in addition to the faulty agents, also the order of message arrival among all agents.
These assumptions are satisfied  by the liveness requirement of $\calCD$.  Liveness ensures that communication is reliable, as any message sent among correct miners is eventually received, yet there is no bound on the finite delay of message arrival, as postulated by the model. Correctness of an agent requires it to be correct in all computations,   hence the system has to be resilient to an adversary that controls not only faulty agents but also the order of transitions by correct agents.  Controlling the order of Receive transitions amounts to control of the order of message arrival, as postulated by the model.
The following proposition is the analogue of Proposition \ref{proposition:GD-local-liveness} for $\calG\calD$, with an analogous proof.

\begin{proposition}[$\calCD$ Grassroots Liveness]\label{proposition:CD-local-liveness}
Let $r$ be a run of $CD$, $p,q \in P$.
If in some configuration $c \in r$, $p$ and $q$ are connected via a friendship path, all of its members follow $q$ in $c$ and are correct in $r$, then for every $q$-block $b$ in $r$ there is a configuration $c' \in r$ for which 
$b \in c'_p$. 
\end{proposition}

\begin{restatable}[]{proposition}{CDgrassroots}\label{proposition:CD-grassroots}
The \CD protocol $\calCD$ is grassroots.
\end{restatable}

\subsection{A Grassroots Implementation of  $\calG\calD$ by $\calCD$}

We use the theorem above to show that:

\begin{theorem}\label{theorem:CD-implements-GD}
$\calCD$ can provide a grassroots implementation of $\calG\calD$.
\end{theorem}

We employ the notion of monotonic-completeness wrt to a partial order to prove the correctness of the implementation:
\begin{definition}[Monotonically-Complete Transition System]\label{definition:monotonic}
A transition system $TS=(P,S(P),T,\lambda)$ over $P\subseteq \Pi$ and a local states function $S$ is \temph{monotonically-complete} wrt a partial order $\prec$ if it is monotonic wrt $\prec$ and  $c0 \xrightarrow{*} c \subseteq T$ and $c \preceq c'$ implies that $c \xrightarrow{*} c' \subseteq T$.
\end{definition}
The partial order $\prec_S$ is often too broad to satisfy this definition, hence in the following we consider a restriction of it, namely a subset $\prec \subseteq \prec_S$.  

%In the following we abbreviate $\prec_S$ and $\prec_{S'}$ as $\prec$ and $\prec'$, respectively.
\begin{definition}[Order-Preserving Implementation]\label{definition:op-implementation}
Let  $TS=(P,S(P),T,\lambda)$ and $TS'=(P,S'(P),T',\lambda')$ be  transition systems over $P$ and local state functions $S$ and $S'$,  monotonically-complete wrt $\prec \subseteq \prec_S$ and $\prec' \subseteq \prec_{S'}$, respectively.
Then an implementation $\sigma : C' \rightarrow C$ of $TS$ by $TS'$ is \temph{order-preserving} wrt $\prec$ and $\prec'$ if:
\begin{enumerate}[partopsep=0pt,topsep=0pt,parsep=0pt]
    \item \temph{Up condition:} 
     $c1' \preceq_{S'} c2'$ implies that $\sigma(c1') \preceq \sigma(c2')$
    \item \temph{Down condition:}  $c0 \xrightarrow{*}c1 \subseteq T$, $ c1 \preceq c2$ implies that there are $c1',c2' \in C'$ such that $c1= \sigma(c1')$, $c2= \sigma(c2')$, $c0' \xrightarrow{* }c1' \subseteq T'$ and $c1' \preceq' c2'$.
\end{enumerate}
\end{definition}

\begin{definition}[$\sigma$: Locally Safe, Productive, Locally Complete]
Given two transition systems $TS=(P,S,T,\lambda)$ and $TS'=(P',S',T',\lambda')$ and an implementation $\sigma: S'(P) \mapsto S(P)$. 
Then $\sigma$ is:
\begin{enumerate}[partopsep=0pt,topsep=0pt,parsep=0pt]
    \item \temph{Locally Safe} if $c0' \xrightarrow{*} c1' \xrightarrow{} c2' \subseteq T'$ implies that  $c0 \xrightarrow{*} c1  \xrightarrow{*} c2 \subseteq T$ for $c1 := \sigma(c1')$ and $c2 := \sigma(c2')$. If $c1 = c2$ then the $T'$ transition  $c1' \xrightarrow{} c2'$ \temph{stutters} $T$.
    
     \item \temph{Productive} if for every $L \in \lambda$ and every correct run $r'$ of $TS'$,
     either $r'$ has a nonempty suffix $r''$ such that $L$ is not enabled in a configuration in $\sigma(r'')$,
     or every suffix $r''$ of $r'$  \temph{activates} $L$, namely $\sigma(r'')$ has an $L$-transition.
    
    \item \temph{Locally Complete} if  $c0 \xrightarrow{*} c1\xrightarrow{} c2 \subseteq T$, implies that
    $c0' \xrightarrow{*} c1' \xrightarrow{*} c2' \subseteq T'$ for some $c1', c2' \in C'$ such that $c1= \sigma(c1')$ and
     $c2 = \sigma(c2')$. 
\end{enumerate}
\end{definition}

\begin{proposition}[$\sigma$ Correct and Complete]\label{proposition:sigma-correct}
If an implementation $\sigma$ is locally safe and productive then it is correct, and if in addition it is locally complete then it is complete.  
\end{proposition}

\begin{theorem}[Correct \& Complete Implementation Among Monotonically-Complete Transition Systems]\label{theorem:sigma-op}
Assume two transition systems $TS=(P,S(P),T,\lambda)$ and $TS'=(P,S'(P),T',\lambda')$, wrt $P$ and local state functions $S$ and $S'$,  monotonically-complete wrt $\prec \subseteq \prec_S$ and $\prec'\subseteq \prec_{S'}$, respectively, and an implementation $\sigma : C' \rightarrow C$  of $TS$ by $TS'$. 
%Furthermore, assume that $\preceq$ is %strict and 
%unbounded. 
If $\sigma$ is order-preserving  wrt $\prec$ and $\prec'$ and productive then it is correct and complete.
\end{theorem}
With this background, we can prove the main Theorem:

\begin{proof}[Proof Outline of Theorem \ref{theorem:CD-implements-GD}]
We define a piecemeal implementation $\sigma$ of  $\calG\calD$ by $\calCD$ (Def. \ref{definition:sigma-CD-GD}), prove that  $\calG\calD$ and $\calCD$ are monotonically-complete (Prop. \ref{proposition:GD-MC},  \ref{proposition:CD-MC})
wrt $\prec_{\calG\calD}$ (Def. \ref{definition:prec-GD}) and $\prec_{\calCD}$ (Def. \ref{definition:prec-CD}), respectively, 
and prove that $\sigma$ is order-preserving (Prop. \ref{proposition:sigma-op-GD-CD}).  Together, these propositions and Theorem \ref{theorem:sigma-op} imply that $\sigma$ is correct.
The implementation $\sigma$ is piecemeal by construction and $\calCD$ is grassroots by Prop. \ref{proposition:CD-grassroots}.  Hence by Def. \ref{definition:grassroots-implementation}, the pair $(\calCD,\sigma)$ constitutes a grassroots implementation of $\calG\calD$, which completes the proof.
\end{proof}

The implementation $\sigma$ defined next is piecemeal. For each local state $c_p=(B_p,Out_p)$, it ignores the block messages $Out_p$, and
maps each block in $B_p$ to a simple block by stripping its hash pointers and adding an index.

\begin{definition}[$\sigma:\calCD \mapsto\calG\calD$]\label{definition:sigma-CD-GD}
Given a $p$-closed blocklace $B$, the \emph{index} of a $p$-block $b \in B$, $\textit{index}(b)$ is the length of the path of self-pointers from $b$ to the initial $p$-block.
The piecemeal implementation $\sigma$  maps each local state $c_p=(B_p,Out_p)$, $p\in P$, in configuration $c$ over $P \subseteq \Pi$, to the local state
$\sigma(c_p) := \{(q,x,\textit{index}(b)) : b=(h^q,H,x) \in B_p \text{ is self-closed and $p$ follows $q$ in $B$}\}$.
\end{definition}

\subsection{Pseudocode Implementation of Grassroots Cordial Dissemination}
The \CD protocol was specified as a family of asynchronous distributed multiagent transition systems $\calCD$ (Def. \ref{definition:protocol-CD}).

\begin{figure}[tp]
\begin{algorithm}[H]
	\caption{\textbf{Grassroots Cordial Dissemination for Asynchrony}\\ Code for agent $p$, including Algorithm \ref{alg:blocklace}}	\label{alg:CD}
	%\small
	%\begin{changemargin}{0cm}{-1.5cm}
	\begin{algorithmic}[1] \scalefont{0.93}
	\alglinenoPop{counter}

\vspace{0.5em}
	\Upon{decision to send $b \in \textit{blocklace}$ to $q$}   \label{alg:BDA-b-to-q}   \Comment{Decision made externally} 
       \State  \textbf{reliably\_send}  $b$ to $q$   \Comment{2. \textbf{Send-$b$-to-$q$}}
	 \EndUpon

\vspace{0.5em}

	\Upon{$\textbf{receive } b$}   \label{alg:BDA-receive}   \Comment{3. \textbf{Receive-$b$}}
	    \State $\textit{blocklace}\gets \textit{blocklace} \cup \{b\}$  \label{alg:BDA-receive}
	 \EndUpon

	   	 \vspace{0.5em}	
     	\Upon{\textbf{eventually}}
      %(\textit{timer})$}
      \label{alg:disseminate-timeout}
	            \Comment{4. \textbf{Cordially-Send-$b$-to-$q$}}
        \For{all $b\in \textit{blocklace}, q \in P :  \textit{friend}(q) \wedge \textit{follows}(q, b.\textit{creator}) \wedge \lnot\textit{agentObserves}(q,b)$}
	    \State \textbf{reliably\_send}  $b$ to  $q$  \label{alg:send-package}
	    \EndFor
 %    	\State $\textbf{reset}(\textit{timer})$ \label{alg:reset-upon-send-package}
	    \EndUpon

		\alglinenoPush{counter}
	\end{algorithmic}

	%\end{changemargin}
% 	\vspace{-5em}
\end{algorithm}
\vspace{-2em}
\end{figure}

\mypara{Asynchrony} Algorithm \ref{alg:CD} presents pseudocode implementation of the protocol for a single agent $p$ for the model of Asynchrony.  The comments indicate which transition rule is realized by what code.
We assume that the agent $p$ can interact with the protocol application that runs on their personal device. In particular, $p$ can specify the payload for a new block or decide to send any block it knows to any other agent.  As according to the model of asynchrony each message sent is eventually received,  we assume the \textbf{reliably\_send} construct to keep a record of sent messages so as not to send the same message to the same agent twice, implementing the requirement $(q,b) \notin Out$ of the Cordially-Send-$b$-to-$q$ transition. The construct \textbf{upon eventually} is also 
geared for the model of asynchrony: It is a `timeless timeout' with no notion of time, only with a guarantee of eventuality.  

The implementation of grassroots dissemination for asynchrony assumes reliable communication, readily implemented by TCP.  However, for grassroots dissemination to reach its full potential it must be designed for mobile agents using unreliable communication, namely UDP.
With TCP, smartphones can establish a connection only with the help of an agreed-upon third party, which undermines grassroots and digital sovereignty, and the connection needs reestablishing upon roaming to a new IP address. 
With UDP, a connection is not needed and incorporating the  sender's current IP address in a block lets recipients (re)transmit blocks to the sender's updated IP address, allowing roaming agents to remain connected to their friends.  
The notion of an ack/nack message on a single channel is well-known.   Here we generalize it a multiple channels, so that an ack/nack message $b$ (i.e., a block in a blocklace) received by agent $p$ from agent $q$ can also serve as ack/nack regarding the communication between $q$ and $q'$:  If $b$ discloses that $q$ is missing a $q'$ block, then $p$ can provide it to $q$ in lieu of $q'$.   The ability of a block in a blocklace to function as a multichannel ack/nack message shines in a UDP implementation, as a received $p$-block $b$ (even if received indirectly, not from $p$) makes redundant the (re)transmission to $p$ of any block observed by $b$.

\begin{figure}[tp]
\begin{algorithm}[H]
	\caption{\textbf{Grassroots Cordial Dissemination for Mobile Agents Over UDP}\\ Code for agent $p$, including Algorithm \ref{alg:blocklace}}	\label{alg:CD-UDP}
	%\small
	%\begin{changemargin}{0cm}{-1.5cm}
	\begin{algorithmic}[1] \scalefont{0.93}
	\alglinenoPop{counter}

	\Statex \textbf{Local variables:}
    \StateX \textit{timer} \Comment{with \textbf{reset} and \textbf{timeout} operations}
    \StateX $\textit{my\_ip} \gets \bot$ \Comment{My IP address and open UDP port, updated externally}
\vspace{0.5em}

     	\Upon{change to \textit{my\_ip}}     \label{alg:disseminate-new-IP}
	            \Comment{Update friends of my new address}
        \State  $b\gets \{\textit{create\_block}((\textit{`ip\_address'},\textit{my\_ip}))\}$
	    \EndUpon

\vspace{0.5em}
	\Upon{decision to send $b \in \textit{blocklace}$ to address \textit{IP}}   \label{alg:BDA-b-to-q}   \Comment{Decision made externally} 
       \State  \textbf{send}  $b$ to \textit{IP}   \Comment{2. \textbf{Send-$b$-to-$q$}}
	 \EndUpon

\vspace{0.5em}

	\Upon{$\textbf{receive } b$}   \label{alg:BDA-receive}   \Comment{3. \textbf{Receive-$b$}}
	  \State  $\textit{blocklace}\gets \textit{blocklace} \cup \{b\}$  \label{alg:BDA-receive}
	 \EndUpon

	   	 \vspace{0.5em}	
     	\Upon{$\textbf{timeout}(\textit{timer})$}     \label{alg:disseminate-timeout}
	            \Comment{4. \textbf{Cordially-Send-$b$-to-$q$}}
        \For{all $b\in \textit{blocklace}, q \in P :  \textit{friend}(q) \wedge \textit{follows}(q, b.\textit{creator}) \wedge \lnot\textit{agentObserves}(q,b)$}
	    \State \textbf{send}  $b$ to  $ip(q)$  \label{alg:send-package}
	    \EndFor
     	\State $\textbf{reset}(\textit{timer})$ \label{alg:reset-upon-send-package}
	    \EndUpon

\vspace{0.5em}
        \Procedure{$\textit{ip}(q)$}{} \Comment{Most-recently known IP address of $q$} 
        \State Let $b$ be the most recent $q$-block in \textit{blocklace} s.t. $b.\textit{payload}= (\textit{`ip\_address'},\textit{IP})$ 
        \State \Return $\textit{IP}$ 
        \EndProcedure

		\alglinenoPush{counter}
	\end{algorithmic}

	%\end{changemargin}
% 	\vspace{-5em}
\end{algorithm}
\vspace{-2em}
\end{figure}

\mypara{Mobile Agents Communicating via UDP} A refinement of the \CD protocol for mobile (address-hopping) agents communicating over an unreliable network, namely smartphones communicating via UDP, is presented as Alg. \ref{alg:CD-UDP}.
Cordial dissemination over UDP exploits the ack/nak information of blocklace blocks to its fullest, by $p$ retransmitting to every friend $q$ every block $b$ that $p$ knows (not only $p$-blocks) and believes that $q$ needs, until $q$ acknowledges knowing $b$.  The sending of unsolicited blocks (Send-$b$-to-$q$) does not guaranteed delivery, and if delivered does not guarantee acknowledgement, unless the block is a friendship offer (with a $(\textit{`follow'},q)$ payload), and the recipient $q$ accept the offer (creating a block with a $(\textit{`follow'},p)$ payload).  Assuming that timeouts are separated by seconds and mobile address changes are independent and are separated by hours,  the probability of two friends hopping together without one successfully informing the other of its new IP address is around $10^{-7}$.
If the two hopping friends have a stationary joint friend, then it is enough that one of the hoppers successfully informs the stationary friend of the address change, for the other hopper  to soon know this new address from their stationary friend.
Under the same assumptions, the probability of a clique of $n$ friends loosing a member due to all hopping simultaneously is around $10^{-3.6*n}$.  Note that such a loss is not terminal --  assuming that friends have redundant ways to communicate (physical meetings, email, SMS, global social media), new addresses can be communicated and the digital friendship restored.  We note that the timer need not be of fixed duration and global  -- an adaptive timer for each friend might prove more efficient.

\section{Applications of \GD Supporting Digital Sovereignty}\label{section:applications}

We describe applications of \GD that support digital sovereignty by not depending on or employing any third-party resources other than the network itself, and thus provide  motivation for implementing the \GD  protocol.

\subsection{Grassroots Twitter-Like Social Networking}\label{section:Twitter-DBD}

Here we illustrate how a Twitter-like grassroots social network can be realised on top of \GD.
Recall that any block $(p,i,x)$ is uniquely identified by its creator $p$ and index $i$, provided the creator is not equivocating.
The Twitter-like realization employs two types of payloads, $\textit{tweet}(x)$, where $x$ is any text string, and
$\textit{respond}(p,i,x)$, which includes the text string $x$ as a response to the $p$-block with index $i$.

With \GD, an agent $p$ will observe any tweet and any response by every agent that $p$ follows.  But, $p$ will not observe responses to tweets of agents that it follows, if $p$ does not follow the respondents.   This behavior could be desirable, if indeed $p$ prefers not to hear these respondents; or undesirable, if $p$ does not know the respondents but would be interested to hear what they have to say.  This can be addressed by adding a third type of payload,
$\textit{echo}(b)$, where $b$ is a block.  With this construct an agent $p$ could echo responses to its own tweets, for the benefit the agents that follow $p$ but not the respondents.  Doing so could induce  an agent to enlarge the set of agents it follows, if it finds agents whose echoed responses are of interest.
This completes the high-level description of how to implement serverless Twitter-like social networking with grassroots dissemination.  

\subsection{Grassroots WhatsApp-Like Social Networking}\label{section:WhatsApp-DBD}

Here we illustrate how a WhatsApp-like grassroots social network can be realised on top of \GD, using the same three payload types, \textit{tweet}, \textit{respond}, and \textit{echo}.
A group is initiated by $p$ creating block $b$ with a special tweet that declares the group formation and invites friends to join the group.  All communications in the group are direct or indirect responses to $b$.  While all group members are friends of $p$, they may not necessarily follow each other.  Therefore,  $p$ echoes all direct or indirect responses to $b$ by members of the group.  This way every group member can follow the group and contribute to it, and a user-interface can make group communication look and feel like a regular messaging group. The group founder $p$ can easily add or remove agents by echoing or ceasing to echo their responses to $b$. 
A group like this is public in that any follower of $p$ can follow the group. If $p$ wishes the group to be private, it can secretly share a new key-pair with all members of the group (as $p$ knows their public key), and have all group communication be encrypted and decipherable only by group members.    If a member leaves an encrypted group, $p$ has to share a new key-pair with the remaining group members.

Another way to ensure privacy is to initiate a separate blocklace for each new group.  This line of thought will be investigated in a separate paper.

\subsection{Grassroots Currencies}\label{section:Twitter-DBD}

Grassroots currencies were introduced in reference~\cite{shapiro2024gc,lewis2023grassroots}, with an implementation by the \AD protocol $\calA\calD$ (Appendiex \ref{appendix-section:protocol-AD}).  However, grassroots currencies do not need all-to-all dissemination: It is enough that dissemination `follows the money'.  This can be achieved by people adhering to the following conservative principle:  Transact only with friends, and only hold coins created by a friend (not the sovereigns include not only people but also legal persons such as banks, corporations, municipalities and states; hence a person its bank may be `friends' in this technical sense and hold each other's coins). 
The key operation of grassroots currencies is coin redemption,
which provides for fungibility, equivocation-exclusion, credit-risk management, coin-trading, chain payments and arbitrage.  With it, an agent $p$ that holds a $q$-coin requests $q$ to redeem this coin for another coin $q$ holds.  Making and serving this request requires bi-directional communication between $p$ and $q$, which can happen by default with \GD if $p$ and $q$ are friends.
Thus, by following this principle, the \GD protocol $\calG\calD$ 
can provide a correct implementation of grassroots currencies.  Transacting with non-friends is also possible through `direct messages' (the Send transition), but without the ability to verify that the other party is following the protocol. Proving this formally requires the entire mathematical machinery of grassroots currencies~\cite{shapiro2024gc,lewis2023grassroots} and is beyond the scope of this paper.

\section{Future Work \& Discussion}\label{section:conclusions}

\mypara{Grassroots dissemination for grassroots currencies} While the \GD protocol $\calG\calD$ is better-suited for  implementing grassroots currencies than the \AD protocol $\calA\calD$, it should be refined to support a genuinely-practical implementation, allowing an agent to:
(\ia)  Follow another agent starting from an arbitrary block (not from its initial block), in particular from the first block recording a transaction between the two agents.  
(\ib) Unfollow an agent, for example when financial relations between the two have ended, permanently or temporarily.
Such extensions to $\calG\calD$ are possible with little additional implementation effort and with some additional mathematical effort for specifying them and proving them correct.

\mypara{Grassroots and the Internet} Formally, the underlying Internet protocol stack is not grassroots, as IP addresses are allocated top-down, with a central entity (IANA) at the top.  Does this mean that striving for grassroots protocols is futile?  We do not think so.  First, the full benefits of grassroots protocols can be reaped as long as IANA  does not abuse IP address allocation and Internet access is not restricted or abused. 
Second, if the local regime  restricts access to servers/services, but does not shut-down the Internet,  grassroots protocols may still operate.  Third, mobile mesh networks or SPANs (Smartphone ad-hoc networks)~\cite{albrecht2021mesh} allow communities to communicate in times of strife (e.g. demonstrations against the regime) without an Internet provider (but current designs are vulnerable~\cite{albrecht2021mesh}).  Extending the UDP-based \CD protocol to a mobile mesh protocol is a goal of our future research.
Achieving it would realize the original vision of Xerox PARC's Bayou~\cite{demers1994bayou} project, of peer-to-peer update upon physical proximity. 

\mypara{Acknowledgements}
Ehud Shapiro is the Incumbent of The Harry Weinrebe Professorial Chair of Computer Science and Biology at the Weizmann Institute.  I thank Nimrod Talmon and Oded Naor for their comments on an earlier version of the manuscript and Idit Keidar for pointing me to related work.

%%
%% Bibliography
%%

%% Please use bibtex, 

\bibliography{bib}

\begin{thebibliography}{10}

\bibitem{albrecht2021mesh}
Martin~R Albrecht, Jorge Blasco, Rikke~Bjerg Jensen, and Lenka Marekov{\'a}.
\newblock Mesh messaging in large-scale protests: Breaking bridgefy.
\newblock In {\em Cryptographers’ Track at the RSA Conference}, pages
  375--398. Springer, 2021.

\bibitem{almeida2024blocklace}
Paulo~Sérgio Almeida and Ehud Shapiro.
\newblock The blocklace: A universal, byzantine fault-tolerant, conflict-free
  replicated data type.
\newblock {\em arXiv preprint arXiv:X.X}, 2024.

\bibitem{benet2014ipfs}
Juan Benet.
\newblock Ipfs-content addressed, versioned, p2p file system.
\newblock {\em arXiv preprint arXiv:1407.3561}, 2014.

\bibitem{buchegger2009peerson}
Sonja Buchegger, Doris Schi{\"o}berg, Le-Hung Vu, and Anwitaman Datta.
\newblock Peerson: P2p social networking: early experiences and insights.
\newblock In {\em Proceedings of the Second ACM EuroSys Workshop on Social
  Network Systems}, pages 46--52, 2009.

\bibitem{buterin2014next}
Vitalik Buterin et~al.
\newblock A next-generation smart contract and decentralized application
  platform.
\newblock {\em white paper}, 3(37), 2014.

\bibitem{castells1983city}
Manuel Castells.
\newblock {\em The city and the grassroots: a cross-cultural theory of urban
  social movements}.
\newblock Number~7. Univ of California Press, 1983.

\bibitem{castro1999practical}
Miguel Castro, Barbara Liskov, et~al.
\newblock Practical byzantine fault tolerance.
\newblock In {\em OsDI}, volume~99, pages 173--186, 1999.

\bibitem{chockler2007constructing}
Gregory Chockler, Roie Melamed, Yoav Tock, and Roman Vitenberg.
\newblock Constructing scalable overlays for pub-sub with many topics.
\newblock In {\em Proceedings of the twenty-sixth annual ACM symposium on
  Principles of distributed computing}, pages 109--118, 2007.

\bibitem{chockler2007spidercast}
Gregory Chockler, Roie Melamed, Yoav Tock, and Roman Vitenberg.
\newblock Spidercast: a scalable interest-aware overlay for topic-based pub/sub
  communication.
\newblock In {\em Proceedings of the 2007 inaugural international conference on
  Distributed event-based systems}, pages 14--25, 2007.

\bibitem{cohen2003incentives}
Bram Cohen.
\newblock Incentives build robustness in bittorrent.
\newblock In {\em Workshop on Economics of Peer-to-Peer systems}, volume~6,
  pages 68--72. Berkeley, CA, USA, 2003.

\bibitem{demers1994bayou}
Alan Demers, Karin Petersen, Mike Spreitzer, Doug Terry, Marvin Theimer, and
  Brent Welch.
\newblock The bayou architecture: Support for data sharing among mobile users.
\newblock In {\em 1994 First Workshop on Mobile Computing Systems and
  Applications}, pages 2--7. IEEE, 1994.

\bibitem{bitcoin-p2p}
Bitcoin Developer.
\newblock P2p network, Retrieved 20222.
\newblock URL: \url{https://developer.bitcoin.org/devguide/p2p_network.html}.

\bibitem{ethereum-bootnodes}
Ethereum Developer.
\newblock go-ethereum/params/bootnodes.go, Retrieved 2023.
\newblock URL:
  \url{https://github.com/ethereum/go-ethereum/blob/master/params/bootnodes.go}.

\bibitem{gencer2018decentralization}
Adem~Efe Gencer, Soumya Basu, Ittay Eyal, Robbert~van Renesse, and Emin~G{\"u}n
  Sirer.
\newblock Decentralization in bitcoin and ethereum networks.
\newblock In {\em International Conference on Financial Cryptography and Data
  Security}, pages 439--457. Springer, 2018.

\bibitem{keidar2021need}
Idit Keidar, Eleftherios Kokoris-Kogias, Oded Naor, and Alexander Spiegelman.
\newblock All you need is dag, 2021.
\newblock \href {http://arxiv.org/abs/2102.08325} {\path{arXiv:2102.08325}}.

\bibitem{keidar2023cordial}
Idit Keidar, Oded Naor, and Ehud Shapiro.
\newblock Cordial miners: A family of simple and efficient consensus protocols
  for every eventuality.
\newblock In {\em 37th International Symposium on Distributed Computing (DISC
  2023)}. LIPICS, 2023.

\bibitem{lewispye2023flash}
Andrew Lewis-Pye, Oded Naor, and Ehud Shapiro.
\newblock Flash: An asynchronous payment system with good-case linear
  communication complexity.
\newblock {\em arXiv preprint arXiv:2305.03567}, 2023.

\bibitem{lewis2023grassroots}
Andrew Lewis-Pye, Oded Naor, and Ehud Shapiro.
\newblock Grassroots flash: A payment system for grassroots cryptocurrencies.
\newblock {\em arXiv preprint arXiv:2309.13191}, 2023.

\bibitem{bitcoin}
Satoshi Nakamoto.
\newblock Bitcoin: A peer-to-peer electronic cash system, 2019.

\bibitem{pastor2015epidemic}
Romualdo Pastor-Satorras, Claudio Castellano, Piet Van~Mieghem, and Alessandro
  Vespignani.
\newblock Epidemic processes in complex networks.
\newblock {\em Reviews of modern physics}, 87(3):925, 2015.

\bibitem{poupko2023thesis}
Ouri Poupko.
\newblock {\em Computational Foundations of Decentralized Internet-enabled
  Governance}.
\newblock PhD thesis, Weizmann Institute of Science, 2023.

\bibitem{raman2019challenges}
Aravindh Raman, Sagar Joglekar, Emiliano~De Cristofaro, Nishanth Sastry, and
  Gareth Tyson.
\newblock Challenges in the decentralised web: The mastodon case.
\newblock In {\em Proceedings of the Internet Measurement Conference}, pages
  217--229, 2019.

\bibitem{setty2012poldercast}
Vinay Setty, Maarten~van Steen, Roman Vitenberg, and Spyros Voulgaris.
\newblock Poldercast: Fast, robust, and scalable architecture for p2p
  topic-based pub/sub.
\newblock In {\em ACM/IFIP/USENIX International Conference on Distributed
  Systems Platforms and Open Distributed Processing}, pages 271--291. Springer,
  2012.

\bibitem{shah2009gossip}
Devavrat Shah et~al.
\newblock Gossip algorithms.
\newblock {\em Foundations and Trends{\textregistered} in Networking},
  3(1):1--125, 2009.

\bibitem{shahaf2020genuine}
Gal Shahaf, Ehud Shapiro, and Nimrod Talmon.
\newblock Genuine personal identifiers and mutual sureties for sybil-resilient
  community growth.
\newblock In {\em Social Informatics: 12th International Conference, SocInfo
  2020, Pisa, Italy, October 6--9, 2020, Proceedings 12}, pages 320--332.
  Springer, 2020.

\bibitem{shapiro2021multiagent}
Ehud Shapiro.
\newblock Multiagent transition systems: Protocol-stack mathematics for
  distributed computing.
\newblock {\em arXiv preprint arXiv:2112.13650}, 2021.

\bibitem{shapiro2023gsn}
Ehud Shapiro.
\newblock Grassroots social networking: Serverless, permissionless protocols
  for twitter/linkedin/whatsapp.
\newblock In {\em OASIS ’23}. Association for Computing Machinery, 2023.
\newblock \href {https://doi.org/10.1145/3599696.3612898}
  {\path{doi:10.1145/3599696.3612898}}.

\bibitem{shapiro2024gc}
Ehud Shapiro.
\newblock Grassroots currencies: Foundations for grassroots digital economies.
\newblock {\em arXiv preprint arXiv:2202.05619}, 2024.

\bibitem{shapiro2011conflict}
Marc Shapiro, Nuno Pregui{\c{c}}a, Carlos Baquero, and Marek Zawirski.
\newblock Conflict-free replicated data types.
\newblock In {\em Symposium on Self-Stabilizing Systems}, pages 386--400.
  Springer, 2011.

\bibitem{shostak1982byzantine}
Robert Shostak, Marshall Pease, and Leslie Lamport.
\newblock The byzantine generals problem.
\newblock {\em ACM Transactions on Programming Languages and Systems},
  4(3):382--401, 1982.

\bibitem{stoica2001chord}
Ion Stoica, Robert Morris, David Karger, M~Frans Kaashoek, and Hari
  Balakrishnan.
\newblock Chord: A scalable peer-to-peer lookup service for internet
  applications.
\newblock {\em ACM SIGCOMM computer communication review}, 31(4):149--160,
  2001.

\bibitem{strom1998gryphon}
Robert Strom, Guruduth Banavar, Tushar Chandra, Marc Kaplan, Kevan Miller,
  Bodhi Mukherjee, Daniel Sturman, and Michael Ward.
\newblock Gryphon: An information flow based approach to message brokering.
\newblock {\em arXiv preprint cs/9810019}, 1998.

\bibitem{tarr2019secure}
Dominic Tarr, Erick Lavoie, Aljoscha Meyer, and Christian Tschudin.
\newblock Secure scuttlebutt: An identity-centric protocol for subjective and
  decentralized applications.
\newblock In {\em Proceedings of the 6th ACM conference on information-centric
  networking}, pages 1--11, 2019.

\bibitem{yin2019hotstuff}
Maofan Yin, Dahlia Malkhi, Michael~K Reiter, Guy~Golan Gueta, and Ittai
  Abraham.
\newblock Hotstuff: Bft consensus with linearity and responsiveness.
\newblock In {\em Proceedings of the 2019 ACM Symposium on Principles of
  Distributed Computing}, pages 347--356, 2019.

\bibitem{zuboff2019age}
Shoshana Zuboff.
\newblock {\em The age of surveillance capitalism: The fight for a human future
  at the new frontier of power: Barack Obama's books of 2019}.
\newblock Profile books, 2019.

\end{thebibliography}

\appendix

\section{Preliminaries: Asynchronous Distributed Multiagent Transition Systems}\label{section:preliminaries}

Here we introduce additional definitions and results regarding asynchronous distributed multiagent transition systems~\cite{shapiro2021multiagent},  needed for the definition and proofs of grassroots protocols. The original reference introduces them in four stages: transition systems; multiagent; distributed; asynchronous, and in addition to proofs it includes examples illustrating the various concepts. Here, we introduce distributed multiagent transition systems at once and simplify other notions by adhering to this special case.

\begin{definition}[Specification; Implementation: Piecemeal, Safe, Live, Correct and Complete]\label{definition:implementation}
Given two transition systems, $TS=(P,S,T,\lambda)$ over $P$ and $S$ (the \temph{specification}) and $TS'=(P,S',T',\lambda')$ over $P$ and $S'$, \temph{an implementation of $TS$ by $TS'$} is a function $\sigma : C' \rightarrow C$ where $\sigma(c0') = c0$, in which case the pair $(TS', \sigma)$ is referred to as  \temph{an implementation of $TS$}. The implementation is \temph{piecemeal} if  $\sigma$ is also defined for local states and satisfies $\sigma(c')_p = \sigma(c'_p)$ for every $c'\in C'$ and $p \in P$.
Given a computation $r'= c'_1\rightarrow c'_2 \rightarrow \ldots$  of $TS'$, $\sigma(r')$ is the (possibly empty) computation obtained from the computation $\sigma(c'_1)\rightarrow \sigma(c'_2) \rightarrow \ldots$ by removing consecutively repetitive elements so that $\sigma(r')$ has no \temph{stutter transitions} of the form $c \rightarrow c$. The implementation $(TS', \sigma)$ of $TS$ is \temph{safe/live/correct} if $\sigma$ maps every safe/live/correct $TS'$ run $r'$ to a safe/live/correct $TS$ run $\sigma(r')$, respectively, 
and is \temph{complete} if every correct run $r$ of $TS$ has a correct run $r'$ of $TS'$ such that $\sigma(r')=r$.
\end{definition}
Note that while an implementation in general is defined for configurations, a piecemeal implementation is defined for local states, which entails its definition for configurations.   Not all implementations presented in~\cite{shapiro2021multiagent} are piecemeal, but, as elaborated below, grassroots and piecemeal go hand-in-hand.
A key property of correct and complete implementations is their transitivity:
\begin{proposition}[Transitivity of Correct \& Complete Implementations]\label{proposition:impelementation-transitivity}
The composition of safe/live/correct/complete   implementations is safe/live/correct/complete, respectively.
\end{proposition}

\section{All-to-All Dissemination is not Grassroots}\label{appendix-section:protocol-AD}

As a strawman, we recall the All-to-All Dissemination protocol $\calA\calD$ from~\cite{shapiro2021multiagent} and argue that it is not grassroots.  
We assume a given \emph{payloads function} $\calX$ that maps each set of agents $P$ to a set of payloads $\calX(P)$.  For example, $\calX$ could map $P$ to all strings signed by members of $P$; or to all messages sent among members of $P$, signed by the sender and encrypted by the public key of the recipient; or to all financial transactions among members of $P$. Remember that here $P$ are not `miners' serving transactions by other agents, but are the full set of agents, all participating the in protocol.
 
\begin{definition}[Simple Block, $SB$, $\prec_{SB}$]
  A \temph{simple block} over $P$ is a triple $(p,i,x) \in P \times \calN \times \calX(P)$.  Such a block is referred to as an \temph{$i$-indexed simple $p$-block with payload $x$}.  The local states function $SB$ maps $P$ to the set of all sets of simple blocks over $P$ and $\calX(P)$.
    The partial order $\prec_{SB}$ is defined  by $c \preceq_{SB} c'$ if
   $c, c'$ are configurations over $P$ and $SB$ and $c_p \subseteq c'_p$ for every $p \in P$.
\end{definition}
Note that  $c \preceq_{SB} c'$  implies that $c\prec_{SB} c'$ if $c_p \subset c'_p$ for some $p \in P$ and $c_q = c'_q$ for every $q \ne p \in P$.

\begin{definition}[$\calA\calD$:  \AD~\cite{shapiro2021multiagent}]\label{definition:AD}
\temph{All-to-all dissemination} $\calA\calD$ is a protocol over $SB$ that for each $P \subseteq \Pi$ has the transition system $AD= (P,SB,T,\lambda)$, with correct transitions $T$ having a $p$-transition $c \rightarrow c' \in T_p$, $c'_p = c_p \cup \{b \}$, $b= (p',i,x)$, for every   $p, p' \in P$,  $i \in \NN$, $x\in \calX(P)$, and either:
\begin{enumerate}[partopsep=0pt,topsep=0pt,parsep=0pt]
    \item \textbf{Create}: $p'=p$, $i = \text{ max } \{ j : (p,j,x') \in c_p\}+1$, or
    \item \textbf{$q$-Sent-$b$}: $p'\ne p$, $b \in c_q \setminus c_p$ for some $q \in P$,
    provided $i=1$ or $c_p$ has an $(i-1)$-indexed simple $p'$-block.  
\end{enumerate}
The liveness condition $\lambda$ places all $p$-transitions with the same label in the same set, for every $p \in P$.
\end{definition}
In other words, every agent $p$ can either add a consecutively-indexed simple $p$-block to its local state or obtain a block $b$ it does not have from another agent, provided $b$ is an initial block or $p$ already has $b$'s preceding  block.
The liveness condition ensures that every correct agent will receive any simple block created by a correct agent, and will eventually create a next block, but it leaves agents the freedom as to which blocks to create.

\begin{definition}[$\prec_{\calA\calD}$]\label{definition:prec-AD}
A configuration $c$ over $P$ and $SB(P)$, $P\subseteq \Pi$, is \temph{consistent} if for every $p$-block $b \in c$, $b \in c_p$, and it is \emph{complete} if for every $i$-indexed $q$-block $b\in c_p$, $c_p$ includes every $i'$-indexed $q$-block $b'$, for every $1\le i' < 1$, $p,q \in P$.   The partial order $\prec_{\calA\calD}$ is defined  by $c \preceq_{\calA\calD} c'$ if
  $c\prec_{SB}c'$ and $c$ and $c'$ are consistent and complete configurations over $P \subseteq \Pi$ and $SB$.
\end{definition}

\begin{proposition}\label{proposition:AD-MC}
$\calA\calD$ is monotonically complete wrt $\prec_{\calA\calD}$.
\end{proposition}
\begin{proof}[Proof of Proposition \ref{proposition:AD-MC}]
The initial configuration is consistent and complete by definition, and both $p$-transitions Create and $q$-Sent-$b$ maintain consistency and completeness by construction,
hence all configurations in a correct run are consistent and complete.
Consider a $p$-transition $c\rightarrow c'$  in a correct computation.
Both Create and $q$-Sent-$b$ increase $c_p$, hence $c\prec_{\calA\calD} c'$ and $\calA\calD$ is monotonic wrt $\prec_{\calA\calD}$.

To show that $\calA\calD$ is monotonically-complete wrt $\prec_{\calA\calD}$, consider two configurations $c\prec_{\calA\calD} c'$ over $P$ and $SB$.  Let $B:= B(c')\setminus B(c)$.  Consider the computation from $c$ that first has a Create $q$-transitions for every  $i$-indexed $q$-block $b \in B$, in increasing order of $i$,  followed by a 
$q$-Sent-$b$ $p$-transition for every $p'$-block $b \in B$, ordered according to the index of $b$ for each $p \ne p' \in P$, in some order of $p$ and $p'$.   It can be verified that such a computation is a correct $\calA\calD$ computation, as the conditions for the the  Create 
transitions are fulfilled due to $c$ being complete and since they are done in creasing index order, and the conditions for $q$-Sent-$b$ transitions with $b$ being a $p'$ block are fulfilled  due to $c$ being complete and since they are done in increasing order of the index of $b$, so that $p$ knows the predecessor of $b$, and since $c'$ is consistent then each sent block has already been created.
\end{proof}

\begin{observation}\label{observation:AD-not-grassroot}
The $\calA\calD$ protocol is not grassroots.
\end{observation}
\begin{proof}[Proof of Observation \ref{observation:AD-not-grassroot}]
A run of a group of agents $P$ executing $AD$, when placed within a larger context $P'$, violates liveness. In a run of $AD(P)$ agents create and receive blocks among themselves.    However, the liveness condition of $AD(P')$ is not satisfied if the agents in $P$ retain the same behavior in a run of $AD(P')$: The reason is that in  an $AD(P')$-run, the liveness condition requires that an agent $p \in P$   eventually receives all blocks created by all agents, including agents in $P' \setminus P$. The transition system $AD(P)$ has correct runs that do not satisfy this condition, which cannot be correct behaviors in $AD(P')/P$.  Hence it is not true that $AD(P) \subseteq AD(P')/P$, and thus $\calA\calD$ is not grassroots.
%\qed    
\end{proof}

\begin{observation}\label{obserbation:AD-interactiv-interfering}
The $\calA\calD$ protocol is asynchronous and interactive but interfering.
\end{observation}
\begin{proof}
Examining $\calA\calD$, it can be verified that it is asynchronous and  interactive, but it is an interfering protocol.  We have argued above that $\calA\calD$ is interactive.
Regarding non-interference safety,  agents in their initial state do not interfere with other agents performing Create or $q$-Sent-$b$ transitions amongst themselves.
However, the $\calA\calD$ liveness condition on $q$-Sent-$b$ requires an agent $p$ to eventually receive every block $b$ created by every other agent.  In particular, in the definition above blocks by agents in $P'\setminus P$.  Hence even if $p$ is live in a run $r$ of $P$, the same behavior of $p$ will not be live if the run is extended by agents in $P'\setminus P$, since $p$ would ignore their blocks, violating liveness.  Therefore the $\calA\calD$ protocol is interfering, and hence the following Theorem \ref{theorem:grassroots} does not apply to it.
%\qed
\end{proof}

\section{Proofs}

\GrassrootsLiveness*
\begin{proof}[Proof of Proposition \ref{proposition:GD-local-liveness}]
Let $r, p, q, c$ be as in the Proposition. 
The proof is by induction on $k$, the length of a friendship path $p_1,p_2,\ldots,p_k$ correct in $r$,  between $q=p_1$ and $p=p_k$ in $c$. Let  $b$ a $q$-block in $c_q \setminus c_p$.  If $k=1$ then $p=q$ and $b \in c_p$.  Assume the proposition holds for $k=n$ and we prove it for $k=n+1$.  By the inductive assumption, $p_n$ eventually knows $b$, namely there is a configuration $c' \in r$ such that $b \in c_{p_n}$, and since $p_n$ is correct by assumption it does not delete $b$ from its local state and hence this holds also for every configuration subsequent to $c'$.  Also by assumption, $p_n$ and $p$ are friends and both follow $q$.  Hence $q$-Disseminates-$b$ is enabled in $c'$ and in any subsequent $r$ configuration in which $b$ is not known by $p$. Since $p$ is live in $r$, then  by the liveness requirement on transitions labeled $p_n$-Disseminates-$b$, eventually such a transition must be taken or all such transitions  be disabled, which can happen if either $p_n$  deletes $b$ it from its local state, or if $p_{n+1}$ already knows $b$.  Since members on the path are correct they do not delete $b$ from their local state, this can only happen if $p_{n+1}=p$ knows $b$ via some other transition. In either case the outcome is that $p$ eventually knows $b$ in $r$. This completes the proof.
\end{proof}

\GDgrassroots*
\begin{proof}[Proof of Proposition \ref{proposition:GD-grassroots}]
We argue somewhat informally that the $\calG\calD$ protocol is asynchronous, interactive and non-interfering, satisfying the condition of  Theorem \ref{theorem:grassroots} for a protocol to be grassroots.
\begin{enumerate}
    \item \textbf{Asynchronous}: The protocol is asynchronous as it is monotonic and examining its two $p$-transitions shows that once a $p$-transition is enabled in a configuration, it remains enabled even if local states of other agents increase.
    \item \textbf{Interactive}: The protocol is interactive since when $P$ is embedded in $P'$, members of $P$ can follow members of $P' \setminus P$ and then receive their blocks, a new behavior not available when $P$ run on their own.  
    \item \textbf{Non-interfering}: The protocol is non-interfering since:
    \begin{enumerate}
        \item \textbf{Safety}: a group $P$ can proceeds with its internal dissemination even if there are agents outside $P$ who do nothing, and
        \item \textbf{Liveness}:  Here the difference between \AD and \GD comes to bear. In a $GD(P')$ run, an agent  $p \in P\subset P'$ may choose not to follow agents in  $P' \setminus P$, in which case the liveness condition of $GD(P')$ does not require $p$ to receive blocks from agents in   $P' \setminus P$.  Hence, if $p$ is live in a run of $GD(P)$, then with the same behavior $p$ is also live in a run of $GD(P')$.
    \end{enumerate}
\end{enumerate}
This completes the proof.
\end{proof}

\begin{proposition}\label{proposition:calb-well-defined}
$\calB(P)$ is well-defined.
\end{proposition}
\begin{proof}[Proof of Proposition \ref{proposition:calb-well-defined}]
Given $P\subseteq \Pi$, we enumerate recursively all elements of $\calB(P)$ and prove that the result is unique.  First, let $\calB$ include all initial blocks of the form $(p,\emptyset,x)$, for any $p \in \PP$ and $x\in \calX$.  
Next, iterate adding to $\calB$ all blocks of the form $(h_p,H,x)$, with $p \in \PP$, $x \in \calX$, and $H$ a finite set of hash pointers to blocks in $\calB$. 
Note that the resulting $\calB$ is acyclic and maximal by construction.  To prove that it is unique, assume that there are two sets $\calB \ne \calB'$ that satisfy our construction, and that wlog $\calB \not\subset \calB'$.  
Let $b=(h_p,H,a)$ be a first block such that $b \in \calB \setminus \calB'$, namely a block for which every block $b'$ pointed to by some $h \in H$ is in $\calB \cap \calB'$. As by assumption all blocks pointed to by $H$ are in $\calB'$, then by construction $\calB'$ includes $b$, a contradiction.
\end{proof}

\begin{observation}[$p$-Closure of $\calCD$]\label{observation:CD-p-closure}
Given a run $r$ of $CD(P)$, if $p\in P$ is correct in $r$ then $c_p$ is $p$-closed for every $c\in r$.
\end{observation}
\begin{proof}[Proof of Observation \ref{observation:CD-p-closure}]
The proof is by induction on the index of $c$ in $r$.
In the initial configuration $c0_p=\emptyset$  and hence closed.
Assume $t= c \rightarrow c' \in r$ and that the observation holds for $c$.  If $t$ is a non-$p$ transition then $c_p = c'_p$. Else one can verify that for each $CD$ $p$-transition, $c'_p$ is $p$-closed by construction.
\end{proof}

\begin{proof}[Proof of Proposition \ref{proposition:CD-local-liveness}]
The proof is the same as the proof of Proposition \ref{proposition:GD-local-liveness}, with two small modifications: (\ia) Instead of the single $p$-transition
$q$-Sent-$b$ of $GD$,  employed in both induction arguments, two transitions are needed:  The $q$-transition $q$-Sends-$b$, which enables the $p$-transition Receive-$b$.  Due to the liveness requirement of CD, Receive-$b$ is eventually taken, resulting in $p$ knowing $b$ as required. (\ib) In CD, the notion of agent knowledge relates to the local blocklace, which is only part of the local state, not to the entire local state as in $GD$.
\end{proof}

\CDgrassroots*
\begin{proof}[Proof of Proposition \ref{proposition:CD-grassroots}]
The proof is similar in structure to the proof of Proposition \ref{proposition:GD-grassroots}, but has additional details.
According to Theorem \ref{theorem:grassroots},
an asynchronous, interactive non-interfering protocol is grassroots.  
Consider  $\emptyset \subset P \subset P' \subseteq \Pi$. 
We argue that the $\calCD$ protocol is:
\begin{enumerate}[leftmargin=*]
    \item \textbf{Monotonic}:  Inspecting the five transitions of the protocol, one can verify that $\calCD$ is monotonic wrt the partial order $\prec_{MSB}$, as they all increase one of the arguments of the local state, except for Input-$b$, which moved $b$ from $In$ to $B$.  The partial order $\prec_{MSB}$ is specifically defined to ensure that this transition is also increasing.
    \item \textbf{Asynchronous}. Inspecting the five transitions of the protocol, one can verify  that $\calCD$ is asynchronous wrt $\prec_{MSB}$.  The transition conditions that are not monotonic wrt the subset relation, namely ``no $q$-block observes $b$'' and ``$p$ does not know that $q$ knows $b$'' refer to the local state of $p$ and hence do not hamper asynchrony, which requires a transition to remain enabled despite a change of state by agents other than $p$.
    
    \item \textbf{Interactive}. Consider agents $p \in P$ and $q \in P'\setminus P$.
    In a $CD(P')$ run, $p$ may receive an Offer-to-Follow from $q$ and  choose to Follow $q$, transitions not available in $CD(P)$.  Hence 
    $\calCD$ is interactive according to Definition \ref{definition:non-interfering}.
    
    \item \textbf{Non-interfering}.  According to Definition \ref{definition:non-interfering}, we have to argue, for every $\emptyset \subset P \subset P' \subseteq \Pi$ with transition systems $TS = (P,\textit{MSB},T,\lambda), TS' = (P',\textit{MSB},T',\lambda') \in \calCD$:
\begin{enumerate}
    \item \textbf{Safety}: That for  every transition $c1 \rightarrow c2 \in T$,
$T'$ includes the transition  $c1' \rightarrow c2'$ for which $c1=c1'/P$, $c2=c2'/P$. 

Consider such a $T$ transition.   It only depends on blocks from agents in $P$.  Hence, by Definition \ref{definition:protocol-CD},  $T'$ includes the transition $c1' \rightarrow c2'$, where the agents of $P' \setminus P$ are in their initial state, namely  $c1_p = c2_p = c0_p$ for every $p \in P' \setminus P$, and
$c1'_p = c2'_p = c0'_p$ for every $p \in P \setminus P'$.
    \item \textbf{Liveness}: That for every agent $p \in P$ and run $r$ of $TS$, if $p$ is live in $r$ then it is also live in every run $r'$ of $TS'$ for which $r'/P = r$.  
    
    Consider such a run $r$ of $TS$ and agent $p$ live in $r$.
    The only  additional transitions enabled for $p$ in a run $r'$ of $TS'$ for which $r'/P = r$, beyond those enabled in $r$, are Follow and Offer-to-Follow, for some $q \in P'\setminus P$.  But such transitions have no liveness requirement, and hence $p$ is also live in $r'$.
\end{enumerate}
\end{enumerate} 
This completes the proof.
\end{proof}

\begin{definition}[Dissemination Consistency]\label{definition:dissemination-consistent}
Let $c$ be a complete and consistent configuration over $P\subseteq \Pi$ and $SB$.
We distinguish between a block $b$ and its \emph{occurrences} in $c$,  denoting the occurrence of a block $b$ in the local state $c_p$ as $b_p$, for any $p \in P$.
A \temph{dependency graph} $DG=(V,E)$ over $c$ is a directed graph over the occurrences $V$ of the blocks in $c$ with directed edged $E$ over $V$. If $b_p \rightarrow b_q \in E$ we say that $b_p$ \temph{depends} on $b_q$.  The graph has edges $E$ as follows:
(\ia) For each non-initial $p$-block $b$ in $c$, the edges in $E$ among all occurrences of $b$ form an inverted spanning tree, with  $b_p$ as its sole sink.  The spanning tree encodes dependencies among the occurrences of a non-initial $p$-block $b$ in a possible dissemination order of $b$ starting with $b_p$, namely the creation of $b$ by $p$.
(\ib) For each edge $b_p\rightarrow b_q \in E$ among two occurrences of a non-initial $i$-indexed $p'$-block $b$, there are also the following three additional edges: Two edges, $b_p \rightarrow (q,1,\bot)_p$ and $b_p \rightarrow (p,1,\bot)_q$, which encode the dependency of the receipt of $b$ by $p$ from $q$ on $p$ and $q$ each knowing the other's initial block, namely on being friends. The third edge is  $b_p \rightarrow b'_p$, to the occurrence in $p$ of the $i-1$-indexed $p'$-block.  A configuration $c$ is \emph{dissemination consistent} if there exists a directed dependency graph $DG$ over $c$ that is acyclic.
\end{definition}
Recall that using a Follow transition, an agent $p$ can `invent' an initial $q$-block and include it in its local state, for any $p,q \in P$. Hence initial blocks have no dependencies.  

For two graphs $G=(V,E)$, $G'=(V',E')$,  $G \subseteq G'$ if $V\subseteq V'$ and $E\subseteq E'$.
\begin{definition}[$\prec_{\calG\calD}$]\label{definition:prec-GD}
The partial order $\prec_{\calG\calD}$ is defined  by $c \preceq_{\calG\calD} c'$ if
  $c\preceq_{SB}c'$ and $c, c'$ are consistent, complete (Def. \ref{definition:prec-AD}) and  dissemination consistent (Def. \ref{definition:dissemination-consistent}) configurations over $P \subseteq \Pi$ and $SB$, with acyclic dependency graphs $DG$, $DG'$, respectively, satisfying $DG \subseteq DG'$.
\end{definition}

\begin{proposition}\label{proposition:GD-MC}
$\calG\calD$ is monotonically complete wrt $\prec_{\calG\calD}$.
\end{proposition}

\begin{proof}[Proof of Proposition \ref{proposition:GD-MC}]
The proof is structurally similar to the proof of Proposition \ref{proposition:AD-MC}, with  modifications to take into account the dependency graph. The initial configuration is consistent, complete and grassroots-consistent  by definition, and all transitions maintain consistency and completeness by construction,
hence all configurations in a correct run are consistent and complete.
We show by induction that all configurations are also dissemination  consistent. Consider a $p$-transition $t= c\rightarrow c' \in T$  in a correct computation, with
$c$ being dissemination-consistent wrt an acyclic dependency-graph $DG=(V,E)$, and construct the graph $DG'=(V',E')$ as follows:  
\begin{enumerate}
    \item \textbf{Create}: If $t$ is a Create $p$-transition of a non-initial $i$-indexed block $b$, then let $e= b'_p \rightarrow b_p$ be an edge from the $p$-occurrence of the $i-1$-indexed $p$-block $b'$ to $b_p$, then $V'=V\cup \{b_p\}$,   $E'=E\cup \{e\}$.
    \item \textbf{Follow}:  If $t$ is a Follow transition with an initial block $b$, then  $V'=V\cup \{b_p\}$,   $E'=E$.
    \item \textbf{$q$-Disseminate-$b$}: If $t$ is a $q$-Disseminate-$b$ transition, with $b$ being an $i$-indexed $p'$-block,  then let $e:= b'_p \rightarrow b_p$ be an edge from the $p$-occurrence of the $i-1$-indexed $p'$-block $b'$ to $b_p$, $e1 := (q,1,\bot)_p \rightarrow b_p$,
    $e2 := (p,1,\bot)_q \rightarrow b_p$,
    then $V'=V\cup \{b_p\}$,   $E'=E\cup \{e, e1, e2\}$.
\end{enumerate}
 It can be verified that in all three cases $DG'$ is a dependency-graph of $c'$, and that if $DG$ is acyclic then so is $DG'$.  In addition, each of Create, Follow, and $q$-Sent-$b$ $p$-transitions increases $c_p$. Hence $c\prec_{\calG\calD} c'$ and $\calG\calD$ is monotonic wrt $\prec_{\calG\calD}$.

To show that $\calG\calD$ is monotonically-complete wrt $\prec_{\calG\calD}$, consider two configurations $c\prec_{\calG\calD} c'$ over $P$ and $SB$.  By Def. \ref{definition:prec-GD} the configurations are consistent, complete and  dissemination-consistent configurations over $P \subseteq \Pi$ and $SB$, with some acyclic dependency graphs $DG=(V,E)$, $DG'=(V',E')$, respectively, satisfying $DG \subseteq DG'$.

Let $B$ be a sequence of the block occurrences in $V'\setminus V$, topologically-sorted in accordance with $DG$.    We argue that there is a one-to-one correspondence between members of $B$ and transitions, and that the dependency graph ensures that each of these transitions is enabled, in order.
Consider the computation from $c\xrightarrow{*}c'$ that has transitions for the block occurrences in $B$, in order, as follows.  For the next block occurrence $b_p \in B$, the computation has the next $p$-transition:
\begin{enumerate}
    \item \textbf{Create}: If $b_p$ is a $p$-block, then Create $b$.
    \item \textbf{Follow}: If $b_p$ is an initial $q$-block, $p\ne q$, then Follow $b$.
    \item \textbf{$q$-Sent-$b$}: If $b_p$ is a non-initial block that depends on $b_q$ in $DG$, then  $q$-Sent-$b$. 
\end{enumerate}
It can be verified that such a computation is a correct $\calG\calD$ computation from $c$ to $c'$,
and is consistent with the dependency graph $DG'$, as the conditions and dependencies for
each of the transitions are guaranteed by ordering them in accordance with the dependence graph, and since the configurations are consistent, complete and dissemination-consistent:
The condition for the  Create transitions are fulfilled due to $c$ being complete and since they are done in creasing index order, according to $DG$. Follow transitions have no dependencies. The conditions and dependencies for $q$-Sent-$b$ transitions are fulfilled since, due to
ordering according to $DG$, the block $b$ occurs in $q$'s local state, the friendship between $p$ and $q$ has been established, and $p$ knows the predecessor of $b$ before the transition takes place due to the dependency in $DG$, as required.
Hence $\calG\calD$ is monotonically-complete wrt $\prec_{\calG\calD}$, which completes the proof.
\end{proof}

A CD configuration records explicitly all the dependencies among the block occurrences in it, with one exception -- if the same block message is sent to the same agent by several agents,
the configuration does not record which of the messages was received.  Hence the cordial dependency graph has some nondeterminism.

\begin{definition}[Cordial Dissemination Consistency]\label{definition:cordial-dissemination-consistent}
Let $c$ be a complete and consistent configuration over $P\subseteq \Pi$ and \textit{MSB}, $c_p=(B_p,Out_p)$ for every $p \in P$.
A \temph{dependency graph} $DG=(V,E)$ over $c$ is a directed graph over $V$, occurrences of blocks and block messages in $c$, with directed edged $E$ over $V$ as follows:
(\ia) For every local state $c_p=(B_p,Out_p)$, and every block message $(q,b)_p \in Out_p$, there is an edge $(q,b)_p  \rightarrow b_p \in E$.
(\ib) For every non-$p$-block $b_p\in B_p$ and for some block message $(p,b)_q \in Out_q$, $q \ne p$, there is an edge $b_p \rightarrow (p,b)_q \in E$.
(\ic) In addition, $E$ has edges that encode the $p$-closure of every $q$-block $b_p=(h^q,H,x) \in B$:
If $p=q$ then for every pointer (and if $p\ne q$ then only for a $q$-pointer) $h \in H$ and block $b'_p\in B$ such that $h=hash(b')$,  $b_p \rightarrow b'_p \in E$.
A configuration $c$ is \emph{cordial-dissemination consistent} if there exists a directed dependency graph $DG$ over $c$ that is acyclic.
\end{definition}

\begin{definition}[$\prec_{\calCD}$]\label{definition:prec-CD}
The partial order $\prec_{\calCD}$ is defined  by $c \preceq_{\calCD} c'$ if
  $c\preceq_{SB}c'$ and $c$ and $c'$ are consistent, complete (Def. \ref{definition:prec-AD}) and  cordial dissemination consistent (Def. \ref{definition:cordial-dissemination-consistent}) configurations over $P \subseteq \Pi$ and $SB$ with acyclic dependency graphs $DG$, $DG'$ respectively, satisfying $DG \subseteq DG'$.
\end{definition}

\begin{proposition}\label{proposition:CD-MC}
$\calCD$ is monotonically complete wrt $\prec_{\calCD}$.
\end{proposition}
\begin{proof}[Proof of Proposition \ref{proposition:CD-MC}]
The proof is structurally similar to the proof of Proposition \ref{proposition:GD-MC}, with  modifications to take into account the differences in the dependency graph and transitions. The initial configuration is consistent, complete and grassroots-dissemination consistent  by definition, and all transitions maintain consistency and completeness by construction,
hence all configurations in a correct run are consistent and complete.
We show by induction that all configurations are also grassroots-dissemination  consistent. Consider a $p$-transition $t= c\rightarrow c' \in T$  in a correct computation, with
$c=(B,Out)$ being grassroots-dissemination consistent wrt an acyclic dependency-graph $DG=(V,E)$, and construct the graph $DG'=(V',E')$, $E \subseteq E'$, as follows: 
\begin{enumerate}
        \item  \textbf{Create}:  If Create adds $b_p=(h^p,H,x)$ to $B$, then $V':=V\cup \{b_p\}$,  and for every pointer  $h \in H$ and block $b'_p$ such that $h=hash(b')$,  $b_p \rightarrow b'_p \in E'$.
        
        \item \textbf{Send-$b$-to-$q$/Offer-to-Follow}: If Send-$b$-to-$q$/Offer-to-Follow adds $(q,b_{q,Out})$ to $Out$,
        then $V':=V\cup \{b\}$ and $(q,b) \rightarrow b_q \in E'$.

        \item \textbf{Receive/Follow}: If Receive/Follow adds $b_p$ to $B$,   then $V':=V\cup \{b_p\}$,  and if there is a self-pointer $h \in H$ and block $b'_p$ such that $h=hash(b')$,  $b_p \rightarrow b'_p \in E'$.
\end{enumerate}

 It can be verified that in all three cases $DG'$ is a dependency-graph of $c'$, and that if $DG$ is acyclic then so is $DG'$.  In addition, each of the transitions increases $c_p$ wrt $\prec_{\calCD}$. Hence $c\prec_{\calCD} c'$ and $\calCD$ is monotonic wrt $\prec_{\calCD}$.

To show that $\calCD$ is monotonically-complete wrt $\prec_{\calCD}$, consider two configurations $c\prec_{\calCD} c'$ over $P$ and \textit{MSB}.  By Def. \ref{definition:prec-CD} the configurations are consistent, complete and  cordial-dissemination consistent configurations over $P \subseteq \Pi$ and \textit{MSB}, with some acyclic dependency graphs $DG=(V,E)$, $DG'=(V',E')$, respectively, satisfying $DG \subseteq DG'$.

Let $\bar{V}$ be a sequence of the block occurrences and block messages in $V'\setminus V$, topologically-sorted in accordance with $DG$.    We argue that there is a one-to-one correspondence between members of $\bar{V}$ and transitions, and that the dependency graph ensures by construction that each of these transitions is enabled, in order. Consider the computation from $c\xrightarrow{*}c'$ that has transitions for the block occurrences in $B$, in order, as follows.  For the next element $v \in \bar{V}$, the computation has the next $p$-transition:
\begin{enumerate}
    \item \textbf{Create}: If $v=b_p$ is a $p$-block, then Create $b$.
    \item \textbf{Offer-to-Follow}: If $v=(q,b)$ and $b$ is an initial block,  then Offer-to-Follow with $(q,b)$.
    \item \textbf{Follow}: If $v=b_p$ is an initial $q$-block, $p\ne q$, then Follow $b$.
    \item \textbf{$q$-Sent-$b$}: If $b_p$ is a non-initial block that depends on $b_q$ in $DG$, then  $q$-Sent-$b$.
         \item \textbf{Receive-$b$}: If $v=b_p$ is a non-initial $q$-block, $q\ne p$,
         then Receive-$b$. 
\end{enumerate}
It can be verified that such a computation is a correct $\calCD$ computation from $c$ to $c'$, and is consistent with the dependency graph $DG'$, as the conditions and dependencies for
enabling each of the transitions in turn are guaranteed by ordering them in accordance with the dependence graph, and since the configurations are consistent, complete and cordial-dissemination consistent.
Hence $\calCD$ is monotonically-complete wrt $\prec_{\calCD}$, which completes the proof.
\end{proof}

\begin{proposition}\label{proposition:sigma-op-GD-CD}
$\sigma$ is an order-preserving implementation of $\calG\calD$  by $\calCD$.
\end{proposition}
\begin{proof}[Proof of Proposition \ref{proposition:sigma-op-GD-CD}]
Let $TS=(P,SB,T,\lambda) \in \calG\calD$ and $TS'=(P,\textit{MSB},T',\lambda')\in  \calCD$ be two transition systems over $P$ and $SB/MSB$, respectively.
According to definition \ref{definition:op-implementation}, we have to prove two conditions.
For the Up condition, we it is easy to see from the definition of $\sigma$ that  $c'_1 \preceqCD  c'_2$ for $TS'$ configurations $c'_1, c'_2$  implies that $\sigma(c'_1) \preceqGD \sigma(c'_2)$, as $\sigma$ maps blocklace blocks to simple blocks and ignores block messages. 

For the Down condition, we construct a $TS'$ representative inverse configuration $\hat{\sigma}(c)$ such that $\hat{\sigma}(\sigma(c))=c$ for any $TS$ configuration $c$.  The mapping $\sigma$ hides two types information: The contents of the output $Out$ of each local state, and the non-self pointers in a block (self-pointers are can be recovered from the block indices of simple blocks).  

In constructing the representative inverse $\hat{\sigma}$ function, we assume that payloads and agent identifiers are lexicographically-ordered, and use this to resolve nondeterminism in the choice of a dependency graph and in its topological sorting.  In the choice of a spanning tree for a block $b$ in the dependency graph of $c$, $\hat{\sigma}(c)$ chooses the first one, lexicographically.  
In the topological sort of the dependency graph, lexicographic ordering of the blocks resolves any nondeterminism in the partial order induced by the directed graph.  The result is that  $\hat{\sigma}(c)$ maps each $TS$ configuration $c$ to a sequence $\bar{B}$ of its block occurrences.  

It then converts the sequence $\bar{B}$ into a $TS'$  configuration $c'$, where $c'_p = (B'_p,Out'_p)$ for every $p \in P$, as follows.
For each simple $p$-block $b_p \in c_p$, $b'_p \in B'_p$, where $b'_p$ has a  pointer to every  $q$-block $b^q \in c_p$ that is the most-recent $q$-block that precedes $b_p$ in $\bar{B}$, for every $q\in P$ (if $b$ is not initial this includes one self-pointer).
For each block $b_p$ that depends on a $p'$-block $b_q$,  $(p,b')\in Out'_q$, where $b'$ is the result of the mapping of $b$.  

The result $c'$ of the mapping is a configuration over $P$ and \textit{MSB}.  It is consistent by construction, since each $q$-block in $c'_p$ depends on it being created by $q$.  It is complete since each block points to its predecessor.  To see that it is grassroots-dissemination consistent, construct the dissemination graph $DG'=(V',E')$  by 
first incorporating in $DG'$ the image of $DG$ under $\hat{\sigma}$, and
then replacing each direct edge $b_p \rightarrow b_q$ by two edges,
$b_p \rightarrow (p,b)_q$ and $(p,b)_q \rightarrow b_q$, and finally adding edges for the pointers among blocks that are part of the $p$-closure of every $p$-block in $B_p$, for every $p \in P$.  The construction is in accordance with the definition of a dependency graph (Def. \ref{definition:prec-CD}) and is acyclic since $DG$ is acyclic and the added edges do not form cycles.  Hence  $c0' \preceqCD c'$.  

To see that $\sigma(\hat{\sigma}(c))=c$, note that each block occurrence in $c$ is mapped one-to-one and and inversely by $\sigma$  by $\hat{\sigma}$, and that  message blocks added by $\hat{\sigma}$ are ignored by $\sigma$.
\qed \end{proof}

To prove  Theorem \ref{theorem:grassroots} we employ the  notion of interleaving:
\begin{definition}[Interleaving]\label{definition:interleaving}
Let $\calF$ be a protocol, $P1, P2 \subset \Pi$, $P1 \cap P2 = \emptyset$. $P:= P1 \cup P2$.  Let
$r1 = c1_0 \rightarrow c1_1 \rightarrow c1_2 \rightarrow \ldots$ be a run of $TS(P1)$,  $r2 = c2_0 \rightarrow c2_1 \rightarrow c2_2 \rightarrow \ldots$ a run of $TS(P2)$.
Then an \temph{interleaving} $r$ of $r1$ and $r2$ is the run of
$TS(P1\cup P2)$, 
$r= c_0 \rightarrow c_1 \rightarrow c_2 \rightarrow \ldots$ satisfying:
\begin{enumerate}
    \item $c_0/P1 = c1_0$ and $c_0/P2 =c2_0)$
    \item For all $i\ge 0$, if $c_i/P1 = c1_j$ and $c_i/P2 = c2_k$ 
   then $c_{i+1}/P1 = c1_{j'}$ and $c_{i+1}/P2 = c2_{k'})$ where either 
$j' = j+1$ and $k' = k$ or
$j' = j$ and $k' = k+1$.
\end{enumerate}
\end{definition}
Note that for any $c_i$ such that  $c_i/P1 = c1_j$ and $c_i/P2 = c2_k$ , $i = j+k$ for all $i\ge 0$.  Note that since the following proof refers explicitly to the difference between the subcommunity ($P$ in Definition \ref{definition:grassroots}) and the supercommunity ($P'$ in the definition),  it uses $P1$ for the subcommunity, $P2$ for the difference, and $P = P1 \cup P2$ for the supercommunity.

\GrassrootsProtocol*
\begin{proof}[Proof of Theorem \ref{theorem:grassroots}]
Let $\calF$ be an asynchronous, interactive, and non-interfering protocol, 
$\emptyset \subset P1 \subset P \subseteq \Pi$, $P2 := P \setminus P1$, 
$r1 = c1_0 \rightarrow c1_1 \ldots$ a correct run of $TS(P1)$,  $r2= c2_0 \rightarrow c2_1 \ldots$ a correct run of $TS(P2)$,  $r=c_0 \rightarrow c_1 \rightarrow \ldots$ the interleaving of $r1$ and $r2$, which is a run of $TS(P)$ by construction.  We argue that $r$ is correct. 

Consider any $p$-transition 
$t= (c \rightarrow c') \in r$ for some $p \in P1$ (else $p \in P2$ and the symmetric argument applies).
Since $c/P1 \rightarrow c'/P1 \in T(P1)$ by construction and  $\calF$ is non-interfering, then by Definition \ref{definition:non-interfering} it follows that the $p$-transition that is similar to $t$, except that agents in $P2$ are in their initial state, is in $T(P)$.  Specifically, the $p$-transition $\hat{c} \rightarrow \hat{c}' \in T(P)$,  defined by $\hat{c}/P1 = c/P1$, $\hat{c}/P2 = c2_0$, $\hat{c}'/P1 = c'/P1$, $\hat{c}'/P2 = c2_0$.
Since $\hat{c} \preceq c$ by monotonicity of $\calF$, $(\hat{c}_p \rightarrow \hat{c}'_p) = (c_p \rightarrow c'_p)$ by construction, and the assumption that $\calF$ is asynchronous together imply that $c \rightarrow c' \in T(P')$. As $c \rightarrow c'$ is a generic transition of $r$, the argument holds for all $r$ transitions, and hence $r$ is safe.

To prove that $r$ is live, recall that $r1$ and $r2$ are correct by assumption, hence the agents of $P1$ are live in $r1$ and the agents of $P2$ are live in $r2$, and together with the liveness condition of non-interference this implies 
that all agents of $P$ are live in $r$, namely $r$ is live.  Hence $r$ is correct
and, since $r1$ and $r2$ were arbitrary,  $TS(P1) \subseteq TS(P)/P1$.   The assumption that $\calF$ is interactive
implies that the inclusion is strict, namely $TS(P1) \subset TS(P)/P1$, concluding that $\calF$ is grassroots.
%\qed
\end{proof}

\end{document}